\documentclass[fleqn,usenatbib]{mnras}

\usepackage{amsmath}	
\usepackage{tikz}
\usepackage{threeparttable}
\usepackage{multirow}
\usepackage[T1]{fontenc}
\usepackage{newtxtext,newtxmath}
\usepackage{ae,aecompl}
\usepackage{placeins}
\usepackage{graphicx}	 
\usepackage{bbding}	
\usepackage{color}
\usepackage{gensymb}
\usepackage{comment}
\usepackage{empheq}
\usepackage{float}
\usetikzlibrary{shapes,arrows,positioning,fit,decorations.pathreplacing,calligraphy}

\DeclareRobustCommand{\VAN}[3]{#2}
\let\VANthebibliography\thebibliography
\def\thebibliography{\DeclareRobustCommand{\VAN}[3]{##3}\VANthebibliography}

\usepackage{scalerel,tikz}
\usetikzlibrary{svg.path}
\definecolor{orcidlogocol}{HTML}{A6CE39}
\tikzset{orcidlogo/.pic={
 \fill[orcidlogocol] svg{M256,128c0,70.7-57.3,128-128,128C57.3,256,0,198.7,0,128C0,57.3,57.3,0,128,0C198.7,0,256,57.3,256,128z};
 \fill[white] svg{M86.3,186.2H70.9V79.1h15.4v48.4V186.2z}
 svg{M108.9,79.1h41.6c39.6,0,57,28.3,57,53.6c0,27.5-21.5,53.6-56.8,53.6h-41.8V79.1z M124.3,172.4h24.5c34.9,0,42.9-26.5,42.9-39.7c0-21.5-13.7-39.7-43.7-39.7h-23.7V172.4z}
 svg{M88.7,56.8c0,5.5-4.5,10.1-10.1,10.1c-5.6,0-10.1-4.6-10.1-10.1c0-5.6,4.5-10.1,10.1-10.1C84.2,46.7,88.7,51.3,88.7,56.8z};
}}
\newcommand\orcidicon[1]{\href{https://orcid.org/#1}{\mbox{\scalerel*{
\begin{tikzpicture}[yscale=-1,transform shape]
\pic{orcidlogo};
\end{tikzpicture}
}{|}}}}

\newcommand{\aref}[1]{\hyperref[#1]{Appendix~\ref{#1}}}

\defcitealias{2018MNRAS.477.2716K}{KBFC18}

\usepackage[normalem]{ulem}

\newcommand{\appropto}{\mathrel{\vcenter{
  \offinterlineskip\halign{\hfil$##$\cr
    \propto\cr\noalign{\kern2pt}\sim\cr\noalign{\kern-2pt}}}}}
    
\title[Piercing through the Galactic ISM with APOGEE]{A path towards constraining the evolution of the interstellar medium and outflows in the Milky Way using APOGEE}

\author[Sharda, Ting and Frankel]{Piyush Sharda$^{\orcidicon{0000-0003-3347-7094}\,1}$\thanks{sharda@strw.leidenuniv.nl (PS)},
Yuan-Sen Ting$^{\orcidicon{0000-0001-5082-9536}\,2,3,4,5}$\thanks{yuan-sen.ting@anu.edu.au (YST)},
and Neige Frankel$^{\orcidicon{0000-0002-6411-8695}\,6,7}$\thanks{neige.frankel@utoronto.ca (NF)}
\\
$^{1}$Leiden Observatory, Leiden University, P.O. Box 9513, 2300 RA Leiden, The Netherlands\\
$^{2}$Research School of Astronomy \& Astrophysics, Australian National University, Canberra 2611, Australia\\
$^{3}$School of Computing, Australian National University, Acton, ACT 2601, Australia\\
$^{4}$Department of Astronomy, The Ohio State University, Columbus, OH 43210, USA \\
$^{5}$Center for Cosmology and AstroParticle Physics, The Ohio State University, Columbus, OH 43210, USA \\
$^{6}$Canadian Institute of Theoretical Astrophysics, University of Toronto, Toronto ON M5S 3H8, Canada \\
$^{7}$Department of Astronomy \& Astrophysics, University of Toronto, Toronto ON M5S 3H4, Canada
}

\date{Accepted 2024 May 28. Received 2024 April 29; in original form 2024 January 24}

\pubyear{2024}

\begin{document}
\label{firstpage}
\pagerange{\pageref{firstpage}--\pageref{lastpage}}
\maketitle

\begin{abstract}

In recent years, the study of the Milky Way has significantly advanced due to extensive spectroscopic surveys of its stars, complemented by astroseismic and astrometric data. However, it remains disjoint from recent advancements in understanding the physics of the Galactic interstellar medium (ISM). This paper introduces a new model for the chemical evolution of the Milky Way that can be constrained on stellar data, because it combines a state-of-the-art ISM model with a Milky Way stellar disc model. Utilizing a dataset of red clump stars from APOGEE, known for their precise ages and metallicities, we concentrate on the last 6 billion years -- a period marked by Milky Way's secular evolution. We examine the oxygen abundance in the low-$\alpha$ disc stars relative to their ages and birth radii, validating or constraining critical ISM parameters that remain largely unexplored in extragalactic observations. The models that successfully reproduce the radius -- metallicity distribution and the age -- metallicity distribution of stars without violating existing ISM observations indicate a need for modest differential oxygen enrichment in Galactic outflows, meaning that the oxygen abundance of outflows is higher than the local ISM abundance, irrespective of outflow mass loading. The models also suggest somewhat elevated ISM gas velocity dispersion levels over the past 6 billion years compared to galaxies of similar mass. The extra turbulence necessary could result from energy from gas accretion onto the Galaxy, supernovae clustering in the ISM, or increased star formation efficiency per freefall time. This work provides a novel approach to constraining the Galactic ISM and outflows, leveraging the detailed insights available from contemporary Milky Way surveys.

\end{abstract}


\section{Introduction} 
\label{s:intro}
Understanding the evolution of galaxies requires studying the relationship between their baryonic components, namely, stars and gas. The connections between these components provide information on the complex interaction between baryons and dark matter, and the galactic lifecyle. This is best studied within our own Galaxy, thanks to several rich, large volume spectroscopic datasets that have collectively mapped more than 10 million stars in the Milky Way \citep{2012RAA....12.1197C,2017AJ....154...94M,2021MNRAS.506..150B,gaiadr2,gaiadr3}. With the help of these datasets, we are now starting to put together the history of our Galaxy on a star by star basis (Galactic archaeology), from in-situ formation of stars in the proto Milky Way \citep[e.g.,][]{2020ApJ...897L..18B,2022arXiv220304980B,2023arXiv230609398S} to accretion events and mergers that dumped stars and gas from external galaxies \citep[e.g.,][]{2018Natur.563...85H,2020ARA&A..58..205H,2020MNRAS.493.3363H}. 

Stars not only encode information on the dynamical evolution of the galaxy, but also preserve signatures of the chemistry of the ISM they were born out of. There has been considerable success in using chemical evolution models that lead to insights on the dynamical history, showing that even with implicit influences taken into account, proper modeling can illuminate the evolution of the Milky Way. However, this is complicated by the fact that stars tend to migrate away from their birth location as they age \citep{2002MNRAS.336..785S,2009MNRAS.399.1145S}, and this migration can be as strong as a few kpc over a few Gyr \citep[e.g.,][]{2010ApJ...722..112M,2012MNRAS.420..913B,2012MNRAS.427..358K,2015MNRAS.446..823M,2018ApJ...865...96F,2022MNRAS.511.5639L}. In the last decade, significant investments have been made in combining chemical evolution models with radial migration models, to characterize the detailed chemo-dynamical assembly history of the Galaxy \citep[e.g.,][]{2009MNRAS.399.1145S,2009MNRAS.396..203S,2013A&A...558A...9M,2014A&A...572A..92M,2015A&A...580A.126K,2015ApJ...802..129S,2020MNRAS.496...80V,2021MNRAS.508.4484J,2023MNRAS.523.3791C}.

However, the prescription for gas-phase abundances used in many of these works is either a parametrization that is not physically motivated, or does not arise from first principles, and therefore cannot reflect the complexity of processes in the interstellar medium (ISM) that sets the gas-phase metallicity. As such, these models cannot be used to explore variations in ISM physics \citep[e.g.,][]{2015MNRAS.449.3479S,2021MNRAS.507.5882S,2021A&A...647A..73S,2022arXiv221204515L,2023MNRAS.tmp.1561R}. Further, models that include a prescription for ISM physics either do not incorporate radial variations, or only solve for the mass and metal balance in the ISM, but not energy  \citep[e.g.,][]{2014A&A...572A..92M,2017ApJ...835..224A,2017ApJ...837..183W,2017MNRAS.472.3637G,2018MNRAS.481.2570G,2021MNRAS.508.4484J}, even though numerous Galactic as well as extragalactic observations have pointed out the role of ISM energetics in modulating star formation and subsequent chemical enrichment (e.g., \citealt{2015ApJ...806L..36S,2015MNRAS.447.4018G,2018MNRAS.477.2716K,2019MNRAS.487.3581F,2020A&A...641A..70B,2022MNRAS.513.6177G,2022ApJ...936..137O,2023arXiv230711595I}; Sharda et al. in preparation). There is very limited work on connecting Galactic ISM physics to Galactic Archaeology, even though chemical evolution is clearly dependent on ISM physics \citep{2023arXiv230609133W}. Similarly, we also lack a bridge between Galactic and extragalactic domains, and the evolutionary history of the Galactic ISM inferred from Milky Way chemical evolution models is not generally validated against extragalactic gas-phase observations that often provide more stringent constraints on ISM physics. 

Complementary to the chemical evolution models developed for the Milky Way, there exists a number of sophisticated ISM evolution models designed to study the gas-phase abundances in external galaxies. These non-parameterized ISM models are generally more descriptive in their treatment of ISM physics, with model parameters mostly constrained by (or, naturally reproduce) extragalactic observations \citep[e.g.,][]{2014MNRAS.438.1552F,2019MNRAS.487.3581F,2018MNRAS.477.2716K,2018MNRAS.475.2236K,2021MNRAS.502.5935S,2023arXiv230315853S,2021MNRAS.503.4474Y,2022MNRAS.513.6177G,2023arXiv231204137S}. However, these models have not been extensively tested on the Milky Way (beyond reproducing the time evolution of the oxygen abundance gradient -- \citealt{2021MNRAS.502.5935S}), despite the fact that it is the only galaxy where we can study chemical evolution on a star by star basis. Therefore, the comprehensive treatment of ISM physics that makes these models state of the art has not been investigated using Milky Way observables. Further, as we show in this work, the Milky Way is the only galaxy that can provide insights into some of the parameters in these ISM models that have no constraints from extragalactic observations. 
As the current and even the upcoming generation of simulations remain far from simulating Milky Way like galaxies in a cosmological context with enough resolution to resolve star formation and feedback (e.g., \citealt{2018ApJ...869...94E,2020ApJ...891....2L,2022MNRAS.514..280B,2023MNRAS.521.5972W,2023ApJ...942...35C}; see however, \citealt{2023arXiv230913115H}), there is vast scope for applying semi-analytical ISM models to reproduce Milky Way observables, and infer the evolutionary history of the Galactic ISM.

In this work, we use a combination of a simple forward model for the Milky Way's disc that includes stellar radial migration, and ISM gas-phase metallicity model to predict the radius -- metallicity and age -- metallicity distributions of stars, which we then compare with the observed distribution of red clump stars from The Apache Point Observatory Galactic Evolution Experiment (APOGEE, \citealt{2014ApJ...790..127B,2017AJ....154...94M}) DR14 \citep{2018ApJS..235...42A}.\footnote{We use APOGEE DR14 instead of the more recent DR17 \citep{2022ApJS..259...35A} to directly incorporate the stellar disc model from \citet{2020ApJ...896...15F}.} Our aim here is only to provide a qualitative assessment of how such a combination of gas + stellar chemo-dynamical modeling can be a powerful tool to constrain the evolutionary history of the ISM of the Milky Way. We organize the rest of the paper as follows: \autoref{s:data} describes the APOGEE data we make use of in this work, \autoref{s:models_overview} gives an overview of the stellar disc and ISM chemical evolution models we use, \autoref{s:models_fiducialmodel} introduces the fiducial (or, the base) model we start the analysis with, \autoref{s:analysis} provides the results of the comparison between the APOGEE data and the models, \autoref{s:discussion} presents a discussion of the key features of our analysis, and \autoref{s:conclusions} concludes our work. For the purpose of this paper, we use Solar metallicity $\rm{Z_{\odot}}$ = 0.0134, corresponding to $12 + \log_{10} \rm{O/H} = 8.69$ \citep{2021A&A...653A.141A}.

\section{Data}
\label{s:data}

We retrieve a sample of 20100 red clump stars from the APOGEE survey that delivered near-infrared spectroscopic data within $R_{\rm{gal}} \sim 5-14\,\rm{kpc}$ from the Galactic centre \citep{2014ApJ...790..127B}. These stars are well known for their precise ages and metallicities \citep{2016ARA&A..54...95G}. Red clump stars were identified by predicting their astroseismic parameters from the spectra with a neural network trained on APOKASC2 data \citep{2018ApJS..239...32P}, and published in a catalog by \cite{2019ApJ...878...21T}. The same method also derived stellar ages. Following \citet[figure 1]{2018ApJ...865...96F}, we only use the sub-sample of these stars which are relatively young (age $< 6\,\rm{Gyr}$, or equivalently, $z \leq 0.6$), have low $\rm{[\alpha/Fe]}$ (where $\alpha$ denotes Type II elements), and are confined within $z_{\rm{gal}} = 1\,\rm{kpc}$ from the midplane. These stars belong to the thin disc at a time when the Galaxy had a relatively secular evolution. Further, simulations find that the CGM of Milky Way analogues virialized $\sim$6.5 Gyr ago, and bursty star formation stopped to allow for quiescence, enabling them to form stable gas discs \citep{2021MNRAS.505..889Y,2023MNRAS.525.2241H,2023arXiv230613125S,2024MNRAS.527.6926M}. Hereon, we simply refer to the low-$\alpha$ disc in the Galaxy as the `disc'. We additionally sub-select stars in the main APOGEE sample, that were drawn randomly from the underlying distribution rather than specific programmes (e.g., clusters) that would not represent the intrinsic Milky Way distribution. We end up with a subsample of 15714 stars that we use in this work, covering a Galactocentric distance between $4-14\,\rm{kpc}$. 

\begin{figure}
\includegraphics[width=\columnwidth]{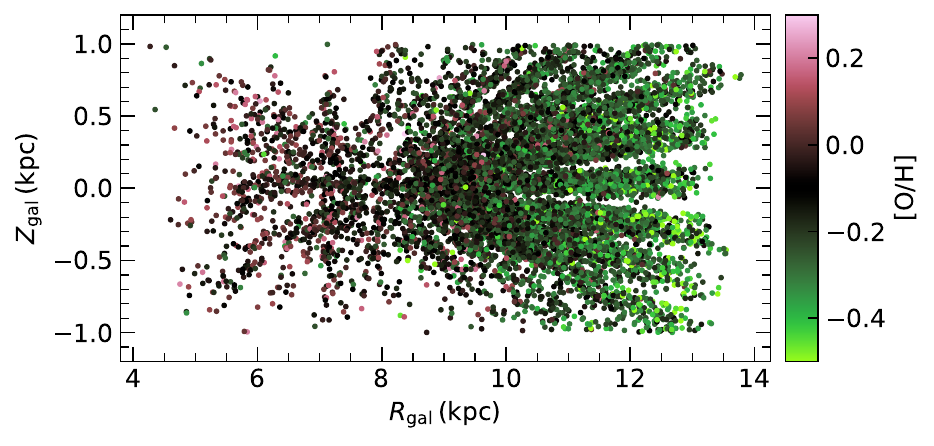}
\caption{Present day location of the subsample of APOGEE red clump giants that we use in this work, color-coded by [O/H].}
\label{fig:apogee_feh_oh}
\end{figure}

\subsection{Metallicity}
\label{s:fehversusoh}
Stellar metallicity is commonly characterised by the iron abundance, whereas ISM metallicity is referenced using oxygen as most of the iron in the ISM is locked in dust \citep[e.g.,][]{2011piim.book.....D}. APOGEE provides both [O/H] and [Fe/H] measurements for the sample we use. Since the ISM model we use is best applicable to $\alpha$ elements, we will use [O/H] for the purpose of our analysis. Additionally, the overall [O/H] measured in the selected sample is expected to reflect the birth abundance (within the uncertainty) because it is not affected by dredge-up \citep{2021arXiv210603912V}. We show the footprint of [O/H] for our subsample in \autoref{fig:apogee_feh_oh}. As \autoref{fig:apogee_feh_oh} shows, the Milky Way has a clear abundance gradient in [O/H], with stars in the inner regions being more metal rich on average (see also, \citealt{2023NatAs...7..951L}). Following \citep{2021arXiv210603912V}, we adopt a conservative uncertainty of $\sigma_{\rm{[O/H]}} = 0.1\,\rm{dex}$ on the oxygen abundance at birth. The uncertainty on stellar ages we adopt is $\sigma_{\tau} = 0.15\,\rm{dex}$ \citep{2020ApJ...896...15F}.



\begin{figure*}
\centering
%
%
\begin{tikzpicture} 

\definecolor{bblue}{RGB}{34,94,168}
\definecolor{ggreen}{RGB}{35,132,67}
\definecolor{rm}{RGB}{140,45,4}

\tikzstyle{main}=[circle, minimum size = 12mm, line width=0.4mm, draw =black!80, node distance = 7mm]
\tikzstyle{ellip}=[ellipse, minimum width=10mm, minimum height=5mm), line width=0.4mm, draw =black!80, node distance = 7mm]
\tikzset{err/.style={
    rectangle,
    minimum size=2mm,
    thick,
    draw=black!100,
    node distance=9mm
}}
\tikzstyle{errc}=[circle, minimum size = 2mm, thick, draw =black!100, node distance = 9mm]

\tikzstyle{connect}=[-latex, thick, ->]
\tikzstyle{box}=[rectangle, draw=black!100]
\tikzstyle{line}=[draw]

   \node[ellip, minimum width=7mm, fill = black!15] (xv) [align=center]{$R_{\mathrm{gal},i}$\\ $z_{\mathrm{gal},i}$};
   \node[ellip, minimum width=25mm,fill = black!15] (feobs) [right=of xv, align=center]{[O/H]$_{\mathrm{obs}\,i}$};
   \node[main, fill = black!15] (tobs) [right=5mm of feobs, label=center:$\tau_{\mathrm{obs}\,i}$] { };

   \node[main, fill=white!100](fe)[above=of feobs, label=center:{[O/H]$_i$}] {};   
   \coordinate (OHnode) at ([xshift=4.5cm]fe.west);

   \node[main, fill=white!100](Lz0)[above=20mm of xv, label=center:{$R$}] {};
   \node[main, fill=white!100](t)[right=37mm of Lz0, label=center:{$\tau_i$}] {};

   \node[ellip, minimum width=28mm, fill = black!15] (evol) [right=of t][draw, align=center] {Evolutionary  \\stage $i$ };

   \node[errc, fill=black!100](Rd0)[above=10mm of Lz0, label=above:structure]{};
   \node[above=2mm of Rd0, label=above:Formation, align=center] {};
 
   \node[errc, fill=black!100](SFH)[right=46.5mm of Rd0, label=above:history]{};
   \node[above=2mm of SFH, label=above:Star formation, align=center] {};

   \node[errc, fill=black!100](secular)[left=35mm of fe, label={[align=center]left: Secular \\evolution}]{}; 

   \node[errc, fill=black!100](errt)[right=54mm of tobs, label=below:$\sigma_{\tau}$] {};
   
   \node[errc, fill=black!100](selfunc)[left=8mm of xv, label={[align=center]left: APOGEE \\spatial selection}]{};
   \node[errc, fill=black!100](RC)[right=55mm of t, label={[align=center]below: Red Clump\\ selection}]{};

   \node[errc, fill=black!100](errfe)[below=10mm of feobs, label=below:$\sigma_{[\mathrm{O}/\mathrm{H}]}$] {}; 

   \node[errc, fill=black!100](pot)[left=22.75mm of errfe, label=left:Potential] {};

   \node[rectangle, fit= (Lz0) (evol) (xv), label={[xshift=46mm]above:{$ i=1,...,N $}}] (rect) {};
   \node[rectangle, fit=(rect), inner sep=4.5mm, draw=black!100] {};

   \node[ellip, fill = white!100, draw=ggreen] (Mg) [below right = 7.7cm and 70mm of secular, align=center, line width=0.8mm] {Mass \\ balance \\ $M_{\rm{g}}$};
   
   \node[ellip, fill=white!100] (radialflows) [above left = 10mm and 20mm of Mg, align=center] {Radial \\ gas flows};
   
   \node[ellip, fill=white!100](starformation)[below=9mm of radialflows][align=center]{Star \\ formation};
   \node[ellip, fill=white!100](starformationparams)[left=5mm of starformation][align=center]{$f_{\rm{sf}},\,\epsilon_{\rm{ff}},\,f_{\rm{gQ}}$ \\ $\sigma_{\rm{SF}},\,C_{\rm{SN}},\,t_{\rm{sf,max}}$};

   \node[ellip, fill=white!100] (radialflowsparams) [above=10mm of starformationparams][align=center] {$\phi_{\rm{nt}},\,F\,(\sigma_{\rm{g}})$};

   \node[ellip, fill=white!100](winds)[below=6mm of starformation][align=center]{Galactic \\ winds};
   \node[ellip, fill=white!100](windsparams)[below=7mm of starformationparams][align=center]{$\phi_{\rm{y}},\,\eta_{\rm{w}},\,\mathcal{Z}_{\rm{w}}$};

   \node[ellip, fill=white!100](accretion)[below=3mm of winds][align=center]{Gas \\ accretion};
   \node[ellip, fill=white!100](accretionparams)[below=19mm of starformationparams][align=center]{$\dot M_{\rm{g,acc}},\,\epsilon_{\rm{in}},\,c$ \\ $\sigma_{\rm{acc}},\,\xi_{\rm{a}}$};
   
   \node[ellip, fill=white!100](diffusion)[above right=0.mm and 5mm of Mg][align=center]{Metal \\ Diffusion};
   \node[main, fill=white!100] (P) [right=6mm of diffusion][draw, align=center] {$\mathcal{P}$};
   \node[main, fill=white!100] (A) [right=55mm of accretion][draw, align=center] {$\mathcal{A}$ };
   \node[main, fill=white!100] (S) [right=55mm of winds][draw, align=center] {$\mathcal{S}$ };

   \node[errc, fill=black!100](tQ)[above left=2mm and 14 mm of radialflows, label=left:Toomre $Q$]{};
   \node[ellip, fill=white!100](toomreQparams)[below=2mm of radialflowsparams, align=center] {$\phi_{\rm{Q}}$};

   \node[ellip, fill=white!100, line width=0.8mm, draw=ggreen](hydrostat)[above = 20 mm of diffusion, align=center]{Hydrostatic \\ equilibrium};
   \node[ellip, fill=white!100](hydrostatparams)[left = 55 mm of hydrostat, align=center]{$\eta,\,f_{\rm{gP}},\,\phi_{\rm{mp}}$};

   \node[errc, fill=black!100](y)[right=21mm of S, label=below:Yield]{};
   \node[below=8mm of y, label=above:$y$, align=center] {};

   \node[ellip, fill=white!100, line width=0.8mm, draw=ggreen](energybalance)[above=6mm of Mg][draw, align=center]{Energy \\ balance \\ $\sigma_{\rm{g}}$};
   \coordinate (energynode) at ([xshift=6.5cm]energybalance.west);

\draw[thick] (OHnode) -- (fe);
\draw[-,thick] (energynode) -- (OHnode);


\draw[decorate,decoration={calligraphic brace,amplitude=10mm,mirror, segment length=0pt}] ([xshift=-9mm, yshift=1mm] hydrostatparams.north west) -- node[midway, left=10mm, font=\large] {ISM} ([xshift=-7mm, yshift=-1mm] accretionparams.south west);

\draw[decorate,decoration={calligraphic brace,amplitude=10mm, segment length=0pt}]([xshift=5mm, yshift=12mm]RC.north east) -- node[midway, right=10mm, font=\large] {Stars} ([xshift=5mm, yshift=-12mm]errt.south east);





  \path     (Rd0) edge [connect] (Lz0)
		(Lz0) edge [connect, bend left=20] (t)
		(SFH) edge [connect] (t)
            (pot) edge [connect] (xv)
		(Lz0) edge [connect] (xv)
		(Lz0) edge [connect] (fe)
		(t) edge [connect, bend right=35] (xv)
		(t) edge [connect] (fe)
		(t) edge [connect] (evol)
		(t) edge [connect] (tobs)
            (secular) edge [connect] (xv)
		(fe) edge [connect] (feobs)
		(errfe) edge [connect] (feobs)
		(errt) edge [connect] (tobs)
		(selfunc) edge [connect] (xv)
		(RC) edge [connect] (evol)
            (y) edge [connect] (S)
            (P) edge [-, thick, bend right=45] (energynode)
            (S) edge [-, thick, bend right=45] (energynode)
            (A) edge [-, thick, bend right=45] (energynode)
		(Mg) edge [connect] (accretion)
		(Mg) edge [connect] (starformation)
		(Mg) edge [connect] (winds)
		(Mg) edge [connect] (radialflows)
		(accretion) edge [connect] (Mg)
		(radialflows) edge [connect] (Mg)
		(starformation) edge [connect] (Mg)
		(winds) edge [connect] (Mg)
		(energybalance) edge [connect] (starformation)
		(energybalance) edge [connect] (diffusion)
		(energybalance) edge [connect] (radialflows)
		(diffusion) edge [connect] (P)
		(diffusion) edge [connect] (S)
		(diffusion) edge [connect, bend right =20] (A)
		(radialflows) edge [connect, bend left=50] (P)
		(accretion) edge [connect] (A)
		(starformation) edge [connect, bend right=40] (S)
		(winds) edge [connect] (S)
		(tQ) edge [connect] (starformation)
		(tQ) edge [connect] (radialflows)
		(hydrostat) edge [connect] (diffusion)
		(hydrostat) edge [connect] (radialflows)
		(starformation) edge [connect] (winds)
		(accretionparams) edge [connect] (accretion)
		(starformationparams) edge [connect] (starformation)
		(radialflowsparams) edge [connect] (radialflows)
		(windsparams) edge [connect] (winds)
		(toomreQparams) edge [connect] (radialflows)
		(toomreQparams) edge [connect] (starformation)
		(hydrostatparams) edge [connect] (hydrostat)
            (Mg) edge [connect] (energybalance);

		
\end{tikzpicture}

\caption{Graphical model that shows the different components of the chemical evolution model, with curly braces marking the corresponding stellar and ISM parts. Filled gray circles denote APOGEE observables, solid dots denote parameters that we keep fixed in this work, and colorless ellipses/circles denote parameters we vary. The green ellipses denote the equilibria we consider in the ISM. All the relevant ISM physics (specified by various symbols in bottom left) gets condensed into three key dimensionless ratios -- $\mathcal{P},\,\mathcal{S}$, and $\mathcal{A}$. We use the model to produce synthetic distributions of stars with a given metallicity, current day galactocentric radius, and age that we then compare against APOGEE data. Definitions of all the parameters are listed in \autoref{tab:taball}.}
\label{fig:graphical_model} 
\end{figure*}
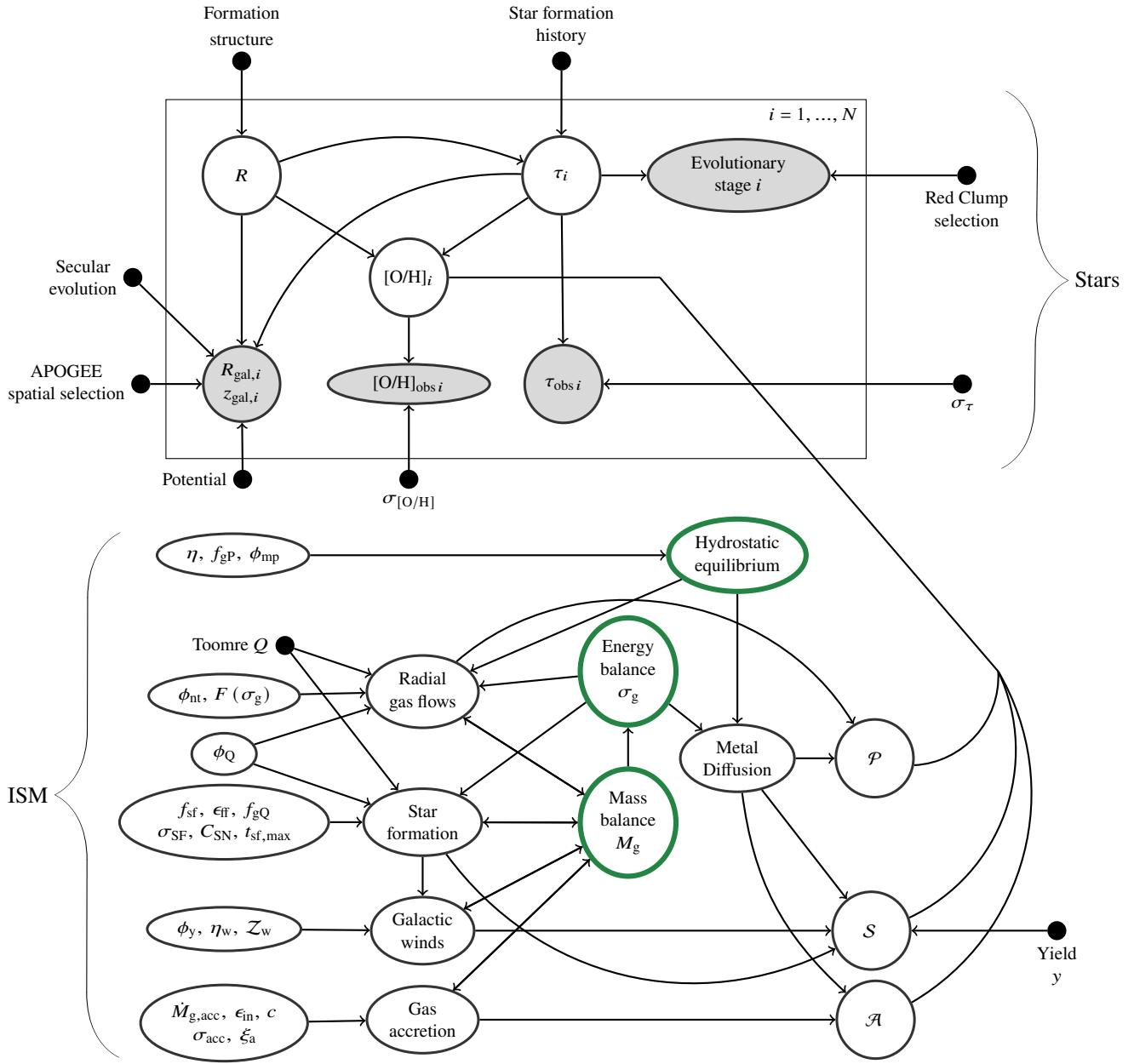

\section{Overview of the models}
\label{s:models_overview}
We aim to compare the predictions from our model to observational data from APOGEE. To this end, we build a forward model, produce mock observations, and compare the distributions of the mock observations to those in the APOGEE dataset.

We describe in brief the stellar disc model and the ISM metallicity model we use in this work. We predict the joint distribution of observables plus the (latent) stellar birth radius $R$ (or Galactocentric radius in the gas) 
\begin{equation}
\begin{split}
p(\mathrm{[O/H]}, \vec{x}, \vec{\mu}, \vec{v}_{los}, \tau, R | \vec{\theta}) &= p(\mathrm{[O/H]}| R, \tau,  \theta_{\mathrm{chem}}) \\ 
& \times p(\vec{x}, \vec{\mu}, \vec{v}_{los}, \tau, R | \theta_{\mathrm{str}}),
\end{split}
\end{equation}
where the first component is the ISM model that we explore in this work, and the second component is the best-fit stellar disc model combined with the APOGEE selection function as in \cite{2020ApJ...896...15F}. Here, [O/H] is the oxygen abundance, $\tau$ is the stellar age, $\vec{v}_{los}$ is the line of sight velocity (from APOGEE DR14), $\vec{x}$ is the 3D spatial position and $\vec{\mu}$ is the proper motion (from Gaia DR2 -- \citealt{gaiadr2}). 

\autoref{fig:graphical_model} shows a graphical model that depicts the overall structure of the modeling we perform in this work. Essentially, the graphical model describes that we keep parameters of the stellar disc model (that is applied to every star) fixed, while exploring the parameter space of the ISM chemical evolution model. We refer the reader to \cite{2018ApJ...865...96F,2020ApJ...896...15F} and \cite{2021MNRAS.502.5935S,2023arXiv230315853S} for a detailed description of the two models, respectively. To produce a synthetic distribution of metallicities, ages and present day radii, we first make a mock stellar dataset and then use the ISM model to paint the abundances based on birth radius and age. We then evaluate the performance of the model against the observed radius -- metallicity and age -- metallicity distributions.

\subsection{ISM metallicity model}
\label{s:models_ismmetallicity}
\cite{2021MNRAS.502.5935S,2023arXiv230315853S} present a first-principles model to explore the physics of radially-resolved gas-phase metallicities in galaxies. In addition to conventional galactic chemical evolution models, this model brings together key metal transport processes like radial gas flows and metal diffusion with gas accretion and outflows to model the metallicity gradients, which makes it applicable to a wide variety of galaxies both in the local Universe and at high-redshift \citep{2021MNRAS.504...53S,2021MNRAS.506.1295S,2022MNRAS.512.3480G,2024arXiv240112319C}. The model incorporates a galactic disc model \citep{2018MNRAS.477.2716K,2022MNRAS.513.6177G} to predict the gas-phase metallicity as a function of the galactocentric radius. The galactic disc model has been validated against numerous gas scaling relations in galaxies \citep[e.g.,][]{Johnson18a,2019MNRAS.486.4463Y,2020MNRAS.495.2265V,2021MNRAS.505.5075Y,2021ApJ...909...12G,2023A&A...673A.153P}.

Before we describe details of the model relevant to this work, it is helpful to provide some remarks about the model. Note that this is an equilibrium model, meaning that the metal distribution is assumed to be in steady-state at a given time. At least for the last 6 Gyr, \cite{2021MNRAS.502.5935S} find that the metal equilibration time is less than the molecular gas depletion time, thus it is safe to assume the galactic disc is in metal equilibrium during this period. 
The model assumes axisymmetry and vertical hydrostatic equilibrium, and therefore does not include azimuthal scatter in metallicity at fixed radius. However, this is not a problem since simulations find that azimuthal scatter only becomes dominant over the radial gradient at $z \gtrsim 1$ in Milky Way-like galaxies, corresponding to lookback times $\gtrsim 8\,\rm{Gyr}$ \citep{2021MNRAS.505.4586B,2022arXiv220303653B}. 
 
At each epoch, the model first solves for mass balance in the disc by equating sources (accretion) and sinks (star formation, transport and radial gas flows, outflows) of gas mass in the disc. Then, we solve for energy balance in the ISM by requiring that the rate of energy injected into the ISM balances the rate of energy dissipation, which sets the gas velocity dispersion. These steps constrain the gas-phase properties of the disc, which we then use to find a solution for metallicity by invoking metal equilibrium. 

The model finds two solutions for the ISM metallicity, depending on the nature of star formation in the galaxy. Specifically, the galaxy evolution model used as an input to the metallicity model \citep{2018MNRAS.477.2716K} separates star formation into the Toomre and giant molecular cloud (GMC) regimes, based on the fact that stars do not form in a continuous volume-filling medium in galaxies like the Milky Way. Instead, most star formation in these galaxies occurs in GMCs with densities $\sim 100 \times$ higher than the midplane density and are hence dynamically decoupled from the disc \citep{2010MNRAS.404.2151C}. For the GMC regime,

\begin{eqnarray}
\lefteqn{\mathrm{[O/H]}(x) = \log_{10}\left(\frac{x\mathcal{S}_{\mathrm{GMC}}}{\mathcal{A}-\mathcal{P}-1} + c_1 x^{\frac{1}{2}\left[\sqrt{\mathcal{P}^2+\,4\mathcal{A}}-\mathcal{P}\right]}\right.}
\nonumber \\
& & \left. + \left(\mathcal{Z}_{R_0} - \frac{\mathcal{S}_{\mathrm{GMC}}}{\mathcal{A}-\mathcal{P}-1} - c_1\right) x^{\frac{1}{2}\left[-\sqrt{\mathcal{P}^2+\,4\mathcal{A}}-\mathcal{P}\right]}\right)\,,
\label{eq:Z_GMC}
\end{eqnarray}
and for the Toomre regime,
\begin{eqnarray}
\lefteqn{\mathrm{[O/H]}(x) = \log_{10}\left(\frac{\mathcal{S}_{\mathrm{Toomre}}}{\mathcal{A}} + c_1 x^{\frac{1}{2}\left[\sqrt{\mathcal{P}^2+\,4\mathcal{A}}-\mathcal{P}\right]}\right.}
\nonumber \\
& & \left. + \left(\mathcal{Z}_{\mathrm{R_0}} - \frac{\mathcal{S}_{\mathrm{Toomre}}}{\mathcal{A}} - c_1\right) x^{\frac{1}{2}\left[-\sqrt{\mathcal{P}^2+\,4\mathcal{A}}-\mathcal{P}\right]}\right)\,,
\label{eq:Z_Toomre}
\end{eqnarray}
where $\mathcal{Z}_{\rm{R_0}}$ is the metallicity at the inner edge of the disc $R_0$, $x = R/R_0$ is the dimensionless radius normalized to the inner edge of the disc, and $c_1$ is a constant of integration that is constrained by the metallicity at the outer edge of the disc (or, the CGM metallicity, $\rm{[O/H]}_{\rm{CGM}}$). We estimate [O/H] in the range $x = 1$ to $x_{\rm{max}}$, where $x_{\rm{max}}$ is the outer edge of the disc. We sample $x_{\rm{max}}$ in the range $13 - 23$, as we describe further in \autoref{s:models_fiducialmodel}.

The dimensionless quantities $\mathcal{P}$, $\mathcal{S}$\footnote{As the subscript in \autoref{eq:Z_GMC} and \autoref{eq:Z_Toomre} implies, the ratio $\mathcal{S}$ is different in the GMC and the Toomre regimes.} and $\mathcal{A}$ respectively describe the ratio of metal advection (due to radial gas flows), metal production and loss (due to star formation and outflows) and gas accretion (due to metal-poor inflows) to metal diffusion due to turbulence. These dimensionless ratios naturally arise in a chemical evolution model that takes into account all major metal transport processes, and includes mass, metal and energy balance in the ISM. We deduce from \autoref{eq:Z_Toomre} and \autoref{eq:Z_GMC} that if $\mathcal{P}$ or $\mathcal{A}$ are larger than $\mathcal{S}$, the resulting metallicities will be lower, with flatter metallicity profiles, and vice-versa. We can intuitively understand this as follows: high levels of star formation or weaker outflows (factors that increase $\mathcal{S}$) will lead to the disc being more enriched as well as retaining more metals, thereby increasing the ISM metallicity. Similarly, higher gas turbulence, transport or accretion (factors that enhance $\mathcal{P}$ and $\mathcal{A}$) will either efficiently mix metals or dilute the overall metallicity, thereby creating flatter metallicity profiles.

For the purpose of this work, we fix $R_0 = 3\,\rm{kpc}$ since 1.) this is the range of the APOGEE data we use, and 2.) the ISM model does not apply to the innermost regions where the bar dominates galactic dynamics. \cite{2021MNRAS.502.5935S} find that for massive local galaxies, $\mathcal{Z}_{\rm{R_0}} \approx \mathcal{S}/\mathcal{A}$. Here, we have also assumed that the rotation curve index is 0 for the range in $x$ we use, since the rotation curve of the Galaxy flattens out around $x \approx 1$ \citep{2016ARA&A..54..529B}.

\begin{table*}
\centering
\caption{List of parameters in the ISM metallicity model. Parameters with $(z)$ vary as a function of redshift. For a full description of these parameters, we refer the reader to \protect\citet{2021MNRAS.502.5935S,2023arXiv230315853S}.}
\begin{tabular}{|l|l|c|c|}
\hline
Parameter & Description & Equation & Fiducial value  \\
\hline
$R$ & Birth radius & ... & ... \\
$R_{\rm{gal}}$ & Present day radius & ... & ... \\
\hline
$R_0$ & Inner edge of the disc & \autoref{eq:Z_GMC} & $3\,\rm{kpc}$ \\
$x$ & Disc radius normalized to $R_0$  & \autoref{eq:Z_GMC} & $1 - x_{\rm{max}}$ \\
$c_{1} (z)$ & Constant of integration in the metallicity equation & \autoref{eq:Z_GMC} & ... \\
$\mathcal{P}\,(z)$ & Ratio of metal advection to metal diffusion$^{1}$ & \autoref{eq:P} & ...\\
$\mathcal{S}\,(z)$ & Ratio of (star formation $\times$ outflow metal enrichment) to metal diffusion$^{1}$ & \autoref{eq:SGMC} & ...\\
$\mathcal{A}\,(z)$ & Ratio of gas accretion to metal diffusion$^{1}$ & \autoref{eq:A} & ...\\
$\mathcal{Z}_{R_0}$ & Metallicity at the inner edge of the disc & \autoref{eq:Z_GMC} & $\mathcal{S}/\mathcal{A}$ \\
\hline
$\eta$ & Scaling factor for rate of turbulent dissipation across the disc scale height$^2$ & \autoref{eq:P} & $1.5$ \\
$\phi_{\rm{Q}} - 1$ & Ratio of gas to stellar Toomre $Q$ parameter$^{2,3,4}$ & \autoref{eq:P} & $2$\\
$\phi_{\rm{nt}}$ & Fraction of gas velocity dispersion due to non-thermal motions$^2$ & \autoref{eq:P} & $1$\\
$f_{\rm{gQ}}\,(z)$ & Effective gas fraction in the disc$^{2,5,6}$ & \autoref{eq:P} & $0 -1$\\
$Q_{\rm{min}}$ & Minimum Toomre $Q$ parameter of stars + gas$^{2,7,8}$ & \autoref{eq:P} & $2.0$\\
$F\,(\rm{\sigma_g})\,(z)$ & Fractional contribution of radial gas flows to turbulence$^{1,9}$ & \autoref{eq:Fsigmag} & $0-1$ \\
$\sigma_{\rm{g}}\,(z)$ & Gas velocity dispersion$^{1,2,9}$ & \autoref{eq:SGMC} & ...\\
$\sigma_{\rm{SF}}\,(z)$ & Turbulence driven by star formation feedback$^2$ & \autoref{eq:Fsigmag} & ...\\
$C_{\rm{SN}}$ & Momentum per unit stellar mass formed injected by clustered supernovae & \autoref{eq:Fsigmag} & 1.0 \\
$\sigma_{\rm{acc}}\,(z)$ & Turbulence driven by accretion$^{6}$ & \autoref{eq:Fsigmag} & ... \\
$\xi_{\rm{a}}\,(z)$ & Accretion-driven turbulence efficiency$^{6,9}$ & \autoref{eq:Fsigmag} & 0.2 \\
$\epsilon_{\mathrm{ff}}$ & Star formation efficiency per free-fall time$^{10,11,12}$ & \autoref{eq:SToomre} & $0.015$\\
$f_{\mathrm{sf}}\,(z)$ & Fraction of star-forming gas$^{10, 13, 14, 15, 16}$ & \autoref{eq:SGMC} & $0-1$\\
$y$ & Total yield of new metals formed in Type II supernovae$^{17}$ & \autoref{eq:SGMC} & 0.028 \\
$Q$ & Toomre $Q$ parameter of stars + gas$^2$ & \autoref{eq:SToomre} & $\geq Q_{\rm{min}}$\\
$f_{\rm{gP}} (z)$ & Effective mid-plane pressure due to self-gravity of gas & \autoref{eq:SGMC} & $f_{\rm{gQ}}$\\
$\phi_{\mathrm{mp}}$ & Ratio of the total to the turbulent pressure at the disc midplane$^2$ & \autoref{eq:SGMC} & $1.4$\\
$v_{\phi}\,(z)$ & Rotational velocity$^{18}$ & \autoref{eq:SGMC} & $220\,\rm{km\,s^{-1}}$\\
$t_{\rm{SF,max}}$ & Maximum gas depletion timescale$^{2,19,20}$ & \autoref{eq:SGMC} & $2\,\rm{Gyr}$\\
$\phi_{\rm{y}}$ & Preferential metal enrichment of galactic outflows$^1$ & \autoref{eq:phiy} & $1.0$\\
$\eta_{\rm{w}}\,(z)$ & Outflow mass loading factor$^{1,21,22,23}$ & \autoref{eq:phiy} & Reference 21\\
$\mathcal{Z}_{\rm{w}}$ & Metallicity of galactic outflows$^1$ & \autoref{eq:phiy} & $1.0$\\
$c\,(z)$ & Halo concentration parameter$^{2,9,24}$ & \autoref{eq:A} & $10$ \\
$\epsilon_{\rm{in}}\,(z)$ & Baryonic accretion efficiency$^{2,9,25}$ & \autoref{eq:A} & $0.1 - 1$\\
$\dot M_{\rm{g,acc}}\,(z)$ & Gas accretion rate$^{2,9,26,27}$ & \autoref{eq:A} & ...\\
\hline
\label{tab:taball}
\end{tabular} \\
\tablenotes{\item References: 1. \citet{2021MNRAS.502.5935S,2023arXiv230315853S}; 2. \citet{2018MNRAS.477.2716K}; 3. \citet{2013MNRAS.433.1389R}; 4. \citet{2016ApJ...827...28G}; 5. \citet{2009ARA&A..47...27K}; 6. \citet{2015ApJ...814...13M}; 7. \citet{2016MNRAS.456.2052I}; 8. \citet{2023arXiv230207823F}; 9. \citet{2022MNRAS.513.6177G}; 10. \citet{2012ApJ...745...69K}; 11. \citet{2018MNRAS.477.4380S}; 12. \citet{2019MNRAS.487.4305S}; 13. \citet{2017ApJS..233...22S}; 14. \citet{2018MNRAS.476..875C}; 15. \citet{2020arXiv200306245T}; 16. \citet{2022arXiv220200690S}; 17. \citet{2019MNRAS.487.3581F}; 18. \citet{2016ARA&A..54..529B}; 19. \citet{2008AJ....136.2782L}; 20. \citet{2022MNRAS.516.3006K}; 21. \citet{2017MNRAS.465.1682H}; 22. \citet{2021MNRAS.508.2979P}; 23. \citet{2020MNRAS.494.3971M}; 24. \citet{2009ApJ...704..137J}; 25. \citet{2011MNRAS.417.2982F}; 26. \citet{2008MNRAS.383..615N}; 27. \citet{2010ApJ...718.1001B}} 
\end{table*}

\subsubsection{Metal advection and radial gas flows}
\label{s:P}
Following \citet[equation 23]{2023arXiv230315853S}, we express $\mathcal{P}$ as the ratio of metal advection to metal diffusion
\begin{equation}
\mathcal{P} = \frac{6\eta\phi^2_{\rm{Q}}\phi^{3/2}_{\rm{nt}} f^2_{\rm{gQ}}}{Q^2_{\rm{min}}} F(\sigma_{\rm{g}})\,.
\label{eq:P}
\end{equation}
We summarise the definitions and fiducial values of all the parameters in \autoref{eq:P} in \autoref{tab:taball}. We discuss one critical parameter, $F\,(\sigma_{\rm{g}})$, in detail below.
The steady-state radial gas flow that we incorporate in the ISM metallicity model comes from the evolution of the gas velocity dispersion due to turbulent dissipation, star formation feedback, gas accretion, and non-axisymmetric torques \citep{2010ApJ...724..895K}. Under energy equilibrium, the radial gas flow rate can then be expressed using $F(\sigma_{\rm{g}})$ as \citep[equation 17]{2023arXiv230315853S}
\begin{equation}
    F(\sigma_{\rm{g}}) = 1 - \frac{\sigma_{\rm{SF}}}{\sigma_{\rm{g}}} - \left(\frac{\sigma_{\rm{acc}}}{\sigma_{\rm{g}}}\right)^3\,,
\label{eq:Fsigmag}
\end{equation}
where $\sigma_{\rm{sf}}$ is the gas velocity dispersion powered by star formation feedback in a galactic disc. The exact formulation of $\sigma_{\rm{sf}}$ is provided in \citet[equation 14]{2023arXiv230315853S}; here, we only highlight that it is linearly proportional to the average momentum injected per unit stellar mass formed by a single supernova. The fiducial value we use is $3000\, C_{\rm{SN}}\,\rm{km\,s^{-1}}$ \citep[e.g.,][]{1998ApJ...500...95T,2015MNRAS.451.2757W,2015ApJ...802...99K}, where $C_{\rm{SN}} > 1$ takes into account the fact that the average momentum scales with the number of supernovae that go off simultaneously in the same patch of the ISM (also known as supernova clustering -- \citealt{2017MNRAS.465.2471G,2022arXiv220810528H}). In the fiducial model, we set $C_{\rm{SN}} = 1$. We will later see the potential importance of supernova clustering to explain the abundance patterns of our APOGEE sample. For Milky Way-like galaxies, $\sigma_{\rm{sf}} \sim 6-15 \,\rm{km\,s^{-1}}$ \citep[e.g.,][]{2009ApJ...704..137J,2011ApJ...743...25K,2018MNRAS.477.2716K,2020MNRAS.495.2265V,2020A&A...641A..70B}. 

Gas accretion not only adds mass to the disc, but also injects energy into the disc, depending on the angular momentum of the gas, clumpiness, and where it joins the disc \citep[e.g.,][]{2010A&A...520A..17K,2020MNRAS.498.2415M,2020A&A...641A..70B,2022arXiv220405344F,2022arXiv221009673J}. $\sigma_{\rm{acc}}$ describes the velocity dispersion due to turbulence driven by gas accretion. \citet[equation 16]{2023arXiv230315853S} provide the exact expression for $\sigma_{\rm{acc}}$; critically, it depends on the efficiency with which accreting gas can convert its kinetic energy into turbulence, $\xi_{\rm{a}}$. To date, this fractional parameter has no constraints from direct observations, and it is not yet known how it varies with redshift, although the general expectation is that it increases with redshift \citep{2022MNRAS.513.6177G}. The fiducial value \cite{2023arXiv230315853S} adopt from \cite{2022MNRAS.513.6177G} is $\xi_{\rm{a}} = 0.2\,(1+z)$. 

In energy equilibrium, the rest of the contribution to $\sigma_{\rm{g}}$, if any, comes from gas transport via radial gas flows. Transport is switched off if star formation feedback and gas accretion are sufficient to drive the turbulence required; in such cases, $F(\sigma_{\rm{g}}) = 0$, and the galaxy can take any $Q \geq Q_{\rm{min}}$. Values of $F(\sigma_{\rm{g}}) < 0$ are not allowed in equilibrium.

\subsubsection{Metal production, star formation and outflows}
\label{s:S}
Now, we consider the dimensionless ratio $\mathcal{S}$ that describes the role of metal production and ejection against diffusion. In the GMC regime it is given by
\begin{equation}
\mathcal{S}_{\mathrm{GMC}} = \frac{3\sqrt{2} f_{\mathrm{sf}} f_{\mathrm{gq}} \phi_{\mathrm{Q}}}{Q} \left(\frac{\phi_{\rm{y}} y}{\mathrm{Z_{\odot}}}\right) \left(\frac{r_0 v_{\phi}}{t_{\mathrm{SF,max}} \sigma^2_{\mathrm{g}}}\right)
\label{eq:SGMC}
\end{equation}
and in the Toomre regime,
\begin{equation}
\mathcal{S}_{\mathrm{Toomre}} = \frac{24 \phi_Q  f^2_{\rm{gQ}} \epsilon_{\rm{ff}} f_{\rm{sf}}}{\pi Q^2 \sqrt{3f_{\rm{gP}} \phi_{\rm{mp}}}}\left(\frac{\phi_{\rm{y}} y}{\mathrm{Z_{\odot}}}\right)\left(\frac{v_{\phi}}{\sigma_{\rm{g}}}\right)^2\,.
\label{eq:SToomre}
\end{equation}
As in \autoref{s:P}, we define the symbols and the corresponding values in \autoref{tab:taball}, and discuss the key parameters below.

There is some subtlety in selecting an appropriate value for $f_{\rm{sf}}$, as $f_{\rm{sf}} \ll 1$ for local spiral galaxies like the Milky Way that lie in the GMC regime \citep[e.g.,][]{2012ApJ...745...69K,2017ApJS..233...22S,2018MNRAS.476..875C}. 
Our fiducial choice is to use observational constraints on the ratio of molecular to molecular + atomic gas in galaxies from \cite{2020arXiv200306245T} and \cite{2022arXiv220200690S}. Later on, we will also explore theoretical models that can parameterize $f_{\rm{sf}}$ in terms of local ISM properties \citep{2009ApJ...693..216K,2013MNRAS.436.2747K}.

The fractional parameter $\phi_{\rm{y}}$ describes the differential metal enrichment of galactic outflows, and is one of the most important parameters but least constrained in the model. $\phi_{\rm{y}}$ takes into account the fact that some fraction of newly produced elements (given by the yield $y$) can be preferentially ejected via galactic outflows, such that the metallicity of galactic outflows can be higher than the ISM metallicity of the galaxy. This can happen due to imperfect metal-mixing, or due to metal-rich ejecta from supernovae directly being entrained in an outflow that dumps metals into the CGM. We express it as \citep[equation 26]{2023arXiv230315853S}
\begin{equation}
\phi_{\rm{y}} = 1 - \eta_{\rm{w}}10^{\mathrm{[O/H]}}\left(\frac{10^{\mathrm{[O/H]}_{\odot}}}{y}\right)\left(\frac{\mathcal{Z}_{\rm{w}}}{10^{\mathrm{[O/H]}}}-1\right)\,,
\label{eq:phiy}
\end{equation}
where $\eta_{\rm{w}}$ is the outflow mass loading factor and $\mathcal{Z}_{\rm{w}}$ is the metallicity of the outflow normalized to Solar. Following \cite{2023arXiv230315853S}, we explore three models that specify $\eta_{\rm{w}}$ as a function of galaxy mass or gas fraction: 1.) from the analytical work of \cite{2017MNRAS.465.1682H}, 2.) from FIRE-2 simulations \citep{2018MNRAS.480..800H} studied by \cite{2021MNRAS.508.2979P}, and 3.) from EAGLE simulations \citep{2015MNRAS.446..521S} studied by \cite{2020MNRAS.494.3971M}. Our fiducial choice is the former.

Despite there being heavy theoretical and observational support \citep[e.g.,][]{1999ApJ...513..142M,2011MNRAS.417.2962P,2014MNRAS.438.1552F,2018ApJ...869...94E,2018MNRAS.481.1690C,2020ApJ...904..152L,2021ApJ...918L..16C,2021MNRAS.504...53S,2023arXiv230907955V}, almost all chemical evolution models for the Milky Way simply assume $\phi_{\rm{y}} = 1$ \textit{across cosmic time}, which corresponds to the outflow metallicity being the same as the ISM metallicity (see \citealt{2023MNRAS.523.3791C} for an exception). We leave it as a free parameter in the fiducial model, but later on we will see how models that best match the data prefer $\phi_{\rm{y}} < 1$.

\subsubsection{Metal dilution and gas accretion}
\label{s:A}
Accretion of metal-poor gas from the cosmic web naturally dilutes the overall metal content in galaxies. We express the last dimensionless ratio, $A$, that describes the ratio of gas accretion to metal diffusion as
\begin{equation}
\mathcal{A} = \frac{3G\phi_{\rm{Q}} \dot M_{\rm{g,acc}}}{2\sigma^3_{\rm{g}} \ln \left(x_{\rm{max}}/x_{\rm{min}}\right)}\,,
\label{eq:A}
\end{equation}
where $\dot M_{\rm{g,acc}}$ is the gas accretion rate onto the disc. $\dot M_{\rm{g,acc}}$ is proportional to the baryonic accretion efficiency of dark matter haloes, $\epsilon_{\rm{in}}$, as well as the halo concentration index $c$. 
We neglect the presence of galactic fountains (whereby a galaxy can re-accrete metal-rich gas that it ejected in a past cycle) due to lack of models that self-consistently include mass, metal and energy exchange between the disc and the CGM via fountains, although recent attempts have been made in this direction \citep[e.g.,][]{2022arXiv221109755P,2022arXiv221105115C}.

\subsubsection{Metal diffusion}
\label{s:diffusion}
Once produced, metals also diffuse into the ISM due to turbulence and thermal/magnetorotational instabilities. Although the existence of metal diffusion has been widely acknowledged \citep{2012ApJ...758...48Y,2015MNRAS.449.2588P,2018MNRAS.475.2236K,2019MNRAS.483.3810R,2020MNRAS.496.1891B}, no Galactic chemical evolution models have incorporated it to date. A key reason for this is the absence of energy balance in these models. The \cite{2023arXiv230315853S} ISM model we use includes a treatment for turbulent metal diffusion. The diffusion coefficient in the model is set by the product of the gas scale height and the velocity dispersion \citep[equation 11]{2013RvMP...85..809K}. We refer the reader to \citet[section 2.2]{2021MNRAS.502.5935S} and \citet[section 1.2]{2023arXiv230315853S} for a detailed discussion of how this is achieved in the model; importantly, we point out that metal diffusion can become dominant over other physical processes in setting metallicity variations in some galaxies.

\begin{figure}
\includegraphics[width=\columnwidth]{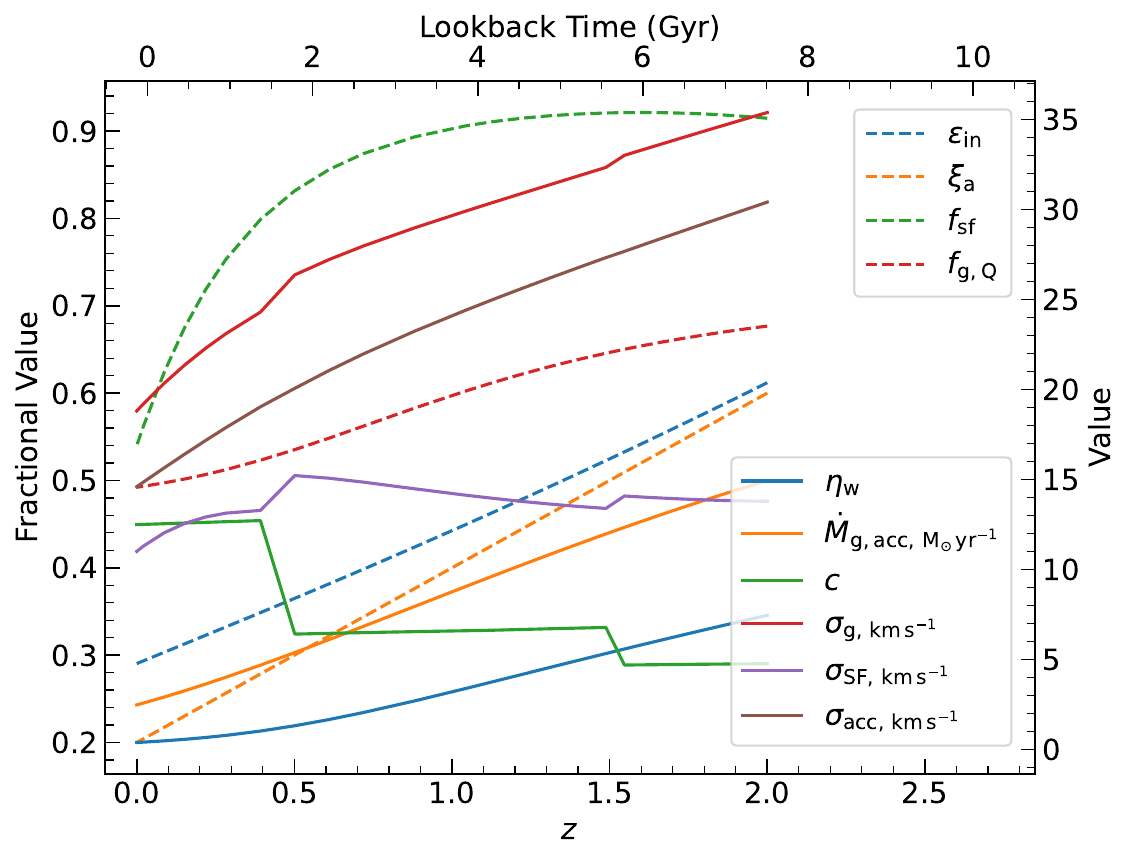}
\caption{Variations in the ISM metallicity model parameters as a function of redshift (or lookback time) in the fiducial model. Left axis shows parameters with fractional values (dashed lines); right axis shows the rest (solid lines). Parameters $\epsilon_{\rm{in}}$ and $\xi_{\rm{a}}$ are used to describe efficiencies with which gas can be accreted and drive turbulence in the disc, respectively. $f_{\rm{sf}}$ and $f_{\rm{gQ}}$ describe gas fraction in the disc. $\eta_{\rm{w}}$ is the mass loading factor, $\dot M_{\rm{g,acc}}$ is the gas accretion rate, and $c$ is the halo concentration index. $\sigma_{\rm{g}}$ is the gas velocity dispersion, a part of which is driven by star formation feedback ($\sigma_{\rm{SF}}$) and gas accretion ($\sigma_{\rm{acc}}$). A detailed description of all the parameters is available in \autoref{tab:taball}.}
\label{fig:fig_allz}
\end{figure}

\subsection{Stellar disc model}
\label{s:models_radialmigration}
We borrow the Milky Way's low-$\alpha$ disc model from \cite{2020ApJ...896...15F}, which predicts the joint distributions $ p(L_z, J_R, \tau, z, R_{\mathrm{gal}} | \vec{\theta}_{\rm{MW}})$. Here, $L_z$ and $J_R$ denote the angular momentum and radial action of the stars.\footnote{We could equally use the guiding radius instead of the Galactocentric radius since the low-$\alpha$ disc model predicts the joint distribution of $L_{z}$ and $R_{\rm{gal}}$ \citep[][figure 6]{2020ApJ...896...15F}, but it would not be more informative for our purpose, since we do not vary the dynamical part of the model.} The model describes where and when stars were born, and how they subsequently changed orbits via radial migration (diffusion in angular momentum) and radial and vertical heating (average increase in radial action and scale-height). These aspects are encoded in structural parameters $\{\vec{\theta}_{\mathrm{str}}\}$. The model also had, in \cite{2020ApJ...896...15F}, a chemical enrichment component for their fitting purpose, with enrichment parameters $ \vec{\theta}_{\mathrm{chem}}$. But in this work, we replace their chemical enrichment by the one in \autoref{s:models_ismmetallicity}.

The model parameters for the Milky Way disc model can be split into structural and chemical evolution parameters $\vec{\theta}_{\rm{MW}} = \{\vec{\theta}_{\mathrm{str}}, \vec{\theta}_{\mathrm{chem}} \}$. We fix $\vec{\theta}_{\mathrm{str}}$ to the best fit of \cite{2020ApJ...896...15F} and in this work we explore $\vec{\theta}_{\mathrm{chem}}$ with the ISM chemical evolution model against red clump stars data. 

Note that the star formation history used by the stellar disc model is not necessarily self-consistent with that produced by the ISM chemical evolution model. For the purpose of this work, we have kept the stellar disc model fixed, so in essence, we treat the two star formation histories independently. However, the star formation rates between the two models are similar \citep{2020ApJ...896...15F}, and satisfy direct constraints from observations, so we expect this inconsistency to not make a significant impact on our inferences. In any case, this inconsistency is present in any work with chemical evolution models that only compare the chemical -- age trends (modelled vs observed) rather than the age distribution of the dataset (predicted by the model vs observed). Here, the star formation history of the stellar disc models is sampled so that the age distribution of the stars in the dataset is well reproduced, but it does not talk to the chemical enrichment part. The star formation history from the chemical evolution model is used only to compare the age--metallicity--radius trends. Ideally, we would like to resolve this inconsistency by fully combining the two models, but this is the scope of a future work.

\section{Fiducial model}
\label{s:models_fiducialmodel}
We begin our analysis by creating a fiducial model. The fiducial model in this context does not mean the model that best describes the data; rather, it is the model where parameters are set to values typical of local spiral galaxies. There is no guarantee that such values also represent the Milky Way. We list the parameter values for the fiducial model in \autoref{tab:taball}. The model has three parameters for which there are no constraints: $\phi_{\rm{y}}, \rm{[O/H]_{CGM}}$, and $x_{\rm{max}}$, the last two of which are only used to create boundary conditions for the model. To statistically sample these parameters for each model we run (for reasons which will become clear in \autoref{s:analysis}), we use Latin Hypercube Sampling (LHS). LHS is a stratified sampling scheme where points sampled for each variable are uniformly distributed but no two samples share the same combination of values. 

To generate an LHS sample, we vary $\phi_{\rm{y}}$ between $0.1-1$, $\rm{[O/H]_{CGM}}$ between $-2$ to $-0.3$, and $x_{\rm{max}}$ between $13-23$, essentially spanning the entire range of $R_{\rm{birth}}$ sampled by the \cite{2020ApJ...896...15F} stellar disc model. We create 10000 samples for each model. As we are interested in learning about the metal entrainment in Galactic outflows, we will later restrict the LHS sampling for $\phi_{\rm{y}}$ to different subsets between 0.1 -- 1.

We begin from $z = 2$, and evolve the model to $z = 0$. We set the initial halo mass at $z=2$, $M_{\rm{h}} = 3.5\times 10^{11}\,\rm{M_{\odot}}$ such that the \cite{2013MNRAS.428.3121M} stellar mass -- halo mass relation gives the present day stellar mass in the disc $M_{\star} \approx 4\times 10^{10}\,\rm{M_{\odot}}$. At each epoch, we solve for the evolution of the gas mass (mass balance), gas velocity dispersion (energy balance), and metallicity (metal balance).\footnote{We use the backward differentiation formula (BDF) routine in \texttt{scipy} to numerically solve for the evolution of these quantities \citep{bdf_ref}.} Note that the exact starting redshift does not matter as the model quickly establishes equilibrium within the first few 100 Myr \citep{2022MNRAS.513.6177G}. We show the evolution of parameters that vary with $z$ in \autoref{fig:fig_allz}. The end product is that we obtain metallicity at each epoch and Galactocentric radius that we pass on to the stellar disc model.

\subsection{Comparison to APOGEE}
\label{s:comparison}
In order to make a statistical comparison between the models and the APOGEE sample, we generate radius -- metallicity and age -- metallicity distributions from the models. We then assess the modeled distributions against the data using simplified versions of four distinct metrics:

\begin{itemize}
    \item \textit{Bhattacharya Distance}: this metric describes the dissimilarity of two distributions, and is sensitive to the shape as well as the overlap between the distributions \citep{Bhattacharya1943}.\footnote{Strictly speaking, the Bhattacharya distance is not a real metric because it does not satisfy the triangle inequality. Similarly, the KL divergence is a distance measure but not a true metric because it is not asymmetric and also does not satisfy the triangle inequality.} In its simplest form, it can be expressed as
    \begin{equation}
    \mathbb{M}_{\mathrm{Bhat}} = \sum_{i=1}^{N_{\rm{bins}}} \sqrt{\mathbb{I}_{\mathrm{data},i} \mathbb{I}_{\mathrm{model},i}} - 1.0\,,
    \end{equation}
    where $\mathbb{I}_{\rm{model}}$ corresponds to the normalized model histogram value (flattened to 1D) in either the radius -- metallicity or the age -- metallicity domains and $N_{\rm{bins}}$ is the total number of bins in these domains. 
    
    \item \textit{KL Divergence}: the Kullback-Leibler distance measure describes the cost of approximating one distribution with another, and is sensitive to the peaks and tails of the distributions. It is expressed as \citep{KLdivergence}
    \begin{equation}
    \mathbb{M}_{\mathrm{KL}} = \sum_{i=1}^{N_{\rm{bins}}} \mathbb{I}_{\mathrm{data},i} |\log_{10}{\mathbb{I}_{\mathrm{data},i} / \mathbb{I}_{\mathrm{model},i}}|
    \end{equation}

    \item \textit{Earth Mover's Distance}: this metric describes the costs associated with transforming one distribution into another. If two distributions are identical, the cost is 0. It is also known as the Wasserstein metric. It is expressed as \citep{emd}
    \begin{equation}
    \mathbb{M}_{\mathrm{EMD}} = \sum_{i=1}^{N_{\rm{bins}}} |\mathbb{I}_{\mathrm{data},i} -\mathbb{I}_{\mathrm{model},i}|
    \end{equation}

    \item \textit{Normalized squared residual}: We use a form of the normalized squared residual as our last metric that can quantify goodness of the fit of the models,
    \begin{equation}
    \mathbb{M}_{\rm{NSR}} = \sum_{i=1}^{N_{\rm{bins}}} \left(\mathbb{I}_{\mathrm{data},i} - \mathbb{I}_{\mathrm{model},i}\right)^2 / \left(\mathbb{I}_{\mathrm{data},i} + \mathbb{I}_{\mathrm{model},i}\right)
    \end{equation}
\end{itemize}
Note that we only apply these metrics if both $\mathbb{I}_{\mathrm{data},i}$ and $\mathbb{I}_{\mathrm{model},i} \neq 0$. We apply a penalty to bins in radius and age where the model has a non-zero value while the APOGEE sample does not, and vice versa. The penalty per bin is equivalent to the square root of the bin value, so that bins where a model deviates more from the APOGEE sample are penalized by a larger value. This penalization plays a crucial role in accurately assessing the performance of different models, as it takes into account their shortcomings as well as their successes. 

We use multiple metrics because they have their own strengths and weaknesses \citep[e.g.,][]{stats1,stats2}. If the model and the data were identical, all the four metrics will yield a value of zero. Therefore, for all the metrics, the best match corresponds to the least value of the metric. These metrics are unavoidably crude, but they are sufficient for our goal to gain a qualitative understanding. A more accurate approach would include fitting the MDF using likelihood-free inference, as we will show in a forthcoming work.

For the remainder of this work, we will use the sum of the four metrics to determine how different models perform,
\begin{equation}
\mathbb{M} = \mathbb{M}_{\mathrm{Bhat}} + 10\mathbb{M}_{\mathrm{KL}} + \mathbb{M}_{\mathrm{EMD}} + \mathbb{M}_{\rm{NSR}}\,,
\label{eq:allmetrics}
\end{equation}
where we have scaled $\mathbb{M}_{\rm{KL}}$ by a factor of 10 to bring its mean magnitude at par with other metrics. The best models are the ones with the least value of the sum, $\mathbb{M}$. In \aref{s:app_diffs}, we show this is reasonable since the four different metrics estimate the performance in similar ways (i.e., it is not the case that one metric predicts a significantly different result than another) but there is some scatter that could otherwise bias the result towards models that excel only on a subset of metrics.

\subsection{Variations in the fiducial model}
\label{s:variations}
As we mention above, the fiducial model well describes the ensemble-averaged evolution of local spiral galaxies, but is not guaranteed or tuned to reproduce Milky Way observables. In this sense, the fiducial model is only a starting point for us to explore the parameter space of Milky Way observations. As our objective is to obtain a qualitative understanding of the gas-phase history of the Milky Way, we focus on varying the ISM model parameters, while keeping the stellar disc model fixed. This approach allows us to determine whether changes in a specific ISM parameter align with the desired outcome in terms of replicating the data. Consequently, it serves as a valuable tool for unraveling the influence of each ISM model parameter on the overall metallicity distribution with radius and age. 

The model consists of two types of parameters: those with numerical values that can be adjusted (e.g., $\epsilon_{\rm{ff}}$), and those that require selecting a specific functional form (e.g., $f_{\rm{sf}}$). For the former, we modify the parameter values by increasing or decreasing them to assess if the adjustments enhance agreement with the data. For the latter, we explore alternative functional forms by substituting the fiducial choice with other existing models or observationally-derived functional forms. We then use the metrics we describe in \autoref{s:comparison} to evaluate the performance of the variation invoked. We list all the models we create in \autoref{tab:tab2}. As we shall see below, the result is insensitive to variations in some parameters, but is clearly influenced by others.

\begin{table*}
\centering
\caption{List of all ISM models created from variations in the fiducial model. Model 1 corresponds to the fiducial model, the parameters of which are listed in \autoref{tab:taball}. Second column lists the modifications applied to the fiducial model. Third column denotes the overall metric score - best models (denoted in bold font) have the lowest score. The typical uncertainty on the combined metric value $\mathbb{M}$ is 0.3. Last column denotes the qualitative inference about ISM physics based on $\mathbb{M}$.}
\begin{tabular}{|l|l|l|l|}
\hline
Model & Parameter(s) varied & $\mathbb{M}$ & Remarks \\
\hline
1 & Fiducial Model (see \autoref{tab:taball}) 
& 
9.95 & \multirow{4}{5cm}{Outflows are somewhat metal enriched \\ as compared to the ISM} \\ 
2 & $0.4 \leq \phi_{\rm{y}} \leq 1.0 $ & 8.76 & \\ 
3 & $0.7 \leq \phi_{\rm{y}} \leq 1.0 $ & 12.43 & \\ 
4 & $0.4 \leq \phi_{\rm{y}} \leq 0.7 $ & 4.11 & \\ 
\hline
5 & $0.4 \leq \phi_{\rm{y}} \leq 0.7 $, $\eta_{\rm{w}}$ from FIRE-2 & 4.83 & \multirow{2}{5cm}{Mass loading is less critical \\ than metal mass loading} \\ 
6 & $0.4 \leq \phi_{\rm{y}} \leq 0.7 $, $\eta_{\rm{w}}$ from EAGLE & 4.67 & \\ 
\hline 
7 & $0.4 \leq \phi_{\rm{y}} \leq 0.7 $, $c(z) \to 0.5\,c(z)$ & 4.77 & \multirow{2}{6cm}{Variations in $c$ not required} \\ 
8 & $0.4 \leq \phi_{\rm{y}} \leq 0.7 $, $c(z) \to 2\,c(z)$ & 7.01 & \\ 
\hline
9 & $0.4 \leq \phi_{\rm{y}} \leq 0.7 $, $\epsilon_{\rm{in}}(z) \to 0.5\,\epsilon_{\rm{in}}(z)$ & 13.16 & \multirow{2}{6cm}{Variations in $\epsilon_{\rm{in}}$ not required} \\ 
10 & $0.4 \leq \phi_{\rm{y}} \leq 0.7 $, $\epsilon_{\rm{in}}(z) \to 2\,\epsilon_{\rm{in}}(z)$ & 12.09 & \\ 
\hline
11 & $0.4 \leq \phi_{\rm{y}} \leq 0.7 $, $f_{\rm{gQ}}(z) \to 0.5\,f_{\rm{gQ}}(z)$ & 35.04 & \multirow{2}{6cm}{Variations in $f_{\rm{gQ}}$ not required} \\ 
12 & $0.4 \leq \phi_{\rm{y}} \leq 0.7 $, $f_{\rm{gQ}}(z) \to 1.5\,f_{\rm{gQ}}(z)$ & 12.94 & \\ 
\hline
13 & $0.4 \leq \phi_{\rm{y}} \leq 0.7 $, $f_{\rm{sf}}\,(z)$ from K13 & 52.33 & Variations in $f_{\rm{sf}}$ not required \\ 
\hline
14 & $0.4 \leq \phi_{\rm{y}} \leq 0.7 $, $\eta \to 2\,\eta$ & 4.83 & Variations in $\eta$ not required \\ 
\hline
15 & $0.4 \leq \phi_{\rm{y}} \leq 0.7 $, $\phi_{\rm{mp}} \to 0.7\,\phi_{\rm{mp}}$ & 4.06 & \multirow{2}{6cm}{Cosmic ray pressure is important in setting hydrostatic equilibrium} \\ 
\textbf{16} & $0.4 \leq \phi_{\rm{y}} \leq 0.7 $, $\phi_{\rm{mp}} \to 1.4\,\phi_{\rm{mp}}$ & \textbf{3.98} & \\ 
\hline
17 & $0.4 \leq \phi_{\rm{y}} \leq 0.7 $, $Q_{\rm{min}} \to 0.5\,Q_{\rm{min}}$ & 17.46 & \multirow{2}{6cm}{Variations in Toomre $Q$ not required} \\ 
18 & $0.4 \leq \phi_{\rm{y}} \leq 0.7 $, $Q_{\rm{min}} \to 1.5\,Q_{\rm{min}}$ & 16.96 & \\ 
\hline
19 & $0.4 \leq \phi_{\rm{y}} \leq 0.7 $, $\phi_{\rm{Q}} \to 2\,\phi_{\rm{Q}}$ & 6.72 & Variations in $\phi_{\rm{Q}}$ not required \\ 
\hline
20 & $0.4 \leq \phi_{\rm{y}} \leq 0.7 $, $\epsilon_{\rm{ff}} \to 0.5\,\epsilon_{\rm{ff}}$ & 5.06 & \multirow{2}{6cm}{Above average star formation efficiency \\ per freefall time} \\ 
\textbf{21} & $0.4 \leq \phi_{\rm{y}} \leq 0.7 $, $\epsilon_{\rm{ff}} \to 2\,\epsilon_{\rm{ff}}$ & \textbf{4.35} & \\ 
\hline
22 & $0.4 \leq \phi_{\rm{y}} \leq 0.7 $, $t_{\rm{sf,max}} \to 0.5\,t_{\rm{sf,max}}$ & 13.18 & \multirow{2}{6cm}{Variations in $t_{\rm{sf,max}}$ not required} \\ 
23 & $0.4 \leq \phi_{\rm{y}} \leq 0.7 $, $t_{\rm{sf,max}} \to 2\,t_{\rm{sf,max}}$ & 23.72 & \\ 
\hline
24 & $0.4 \leq \phi_{\rm{y}} \leq 0.7 $, No GMC & 43.76 & GMC regime of star formation is crucial \\ 
\hline
\textbf{25} & $0.4 \leq \phi_{\rm{y}} \leq 0.7 $, $C_{\rm{SN}} \to 2\,C_{\rm{SN}}$ & \textbf{4.19} & \multirow{2}{6cm}{Supernovae clustering is needed for energy balance} \\ 
26 & $0.4 \leq \phi_{\rm{y}} \leq 0.7 $, $C_{\rm{SN}} \to 5\,C_{\rm{SN}}$ & 3.46 & \\ 
\hline
27 & $0.4 \leq \phi_{\rm{y}} \leq 0.7 $, $\xi_{\rm{a}}(z) \to 0.5\,\xi_{\rm{a}}(z)$ & 4.32 & \multirow{2}{6cm}{Higher accretion induced turbulence \\ is needed for energy balance} \\ 
\textbf{28} & $0.4 \leq \phi_{\rm{y}} \leq 0.7 $, $\xi_{\rm{a}}(z) \to 2\,\xi_{\rm{a}}(z)$ & \textbf{3.94} & \\ 
\hline
\end{tabular}
\label{tab:tab2}
\end{table*}

\section{Physical Constraints on ISM parameters}
\label{s:analysis}

\begin{figure}
\includegraphics[width=\columnwidth]{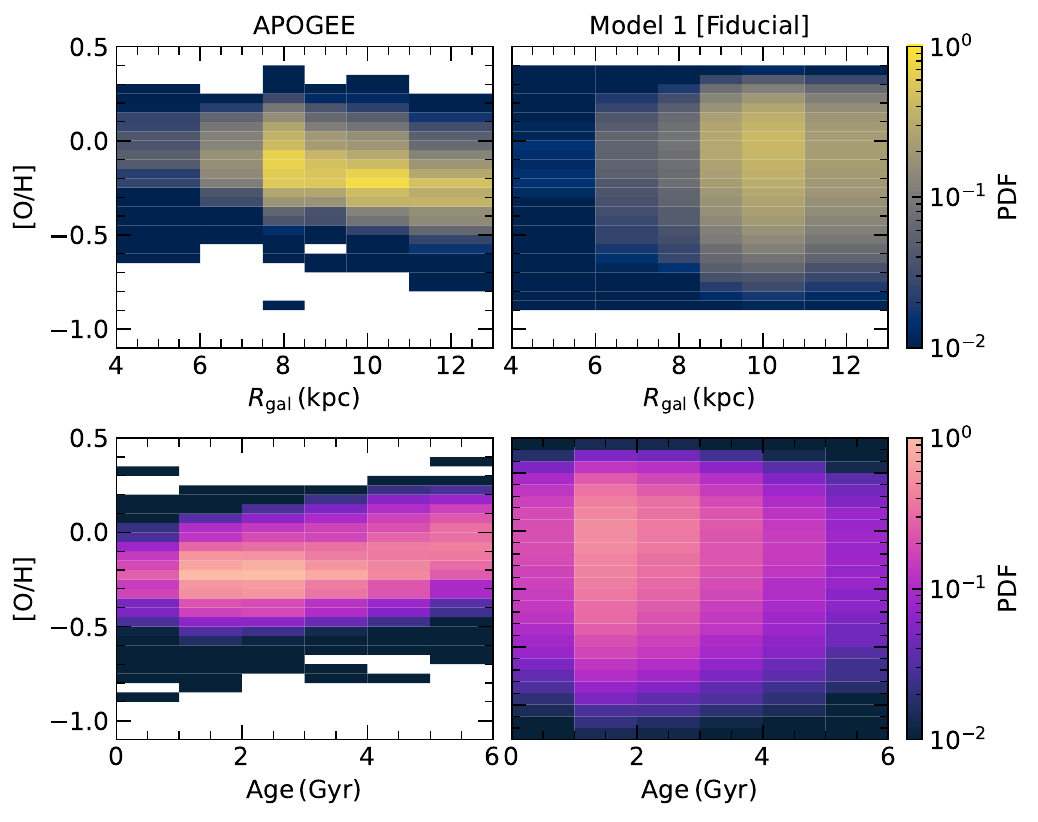}
\caption{Radius -- metallicity and age -- metallicity distributions for the sample of red clump stars from APOGEE DR17 used in this work (left panels), compared against the fiducial model presented in this work. Fiducial model in this context does not mean the model that best describes the data -- it is the starting point for all models. The fiducial values of all ISM parameters are taken from a collection of observations and theoretical calculations, and are listed in \autoref{tab:taball}.}
\label{fig:fig3}
\end{figure}

\begin{figure*}
\includegraphics[width=\columnwidth]{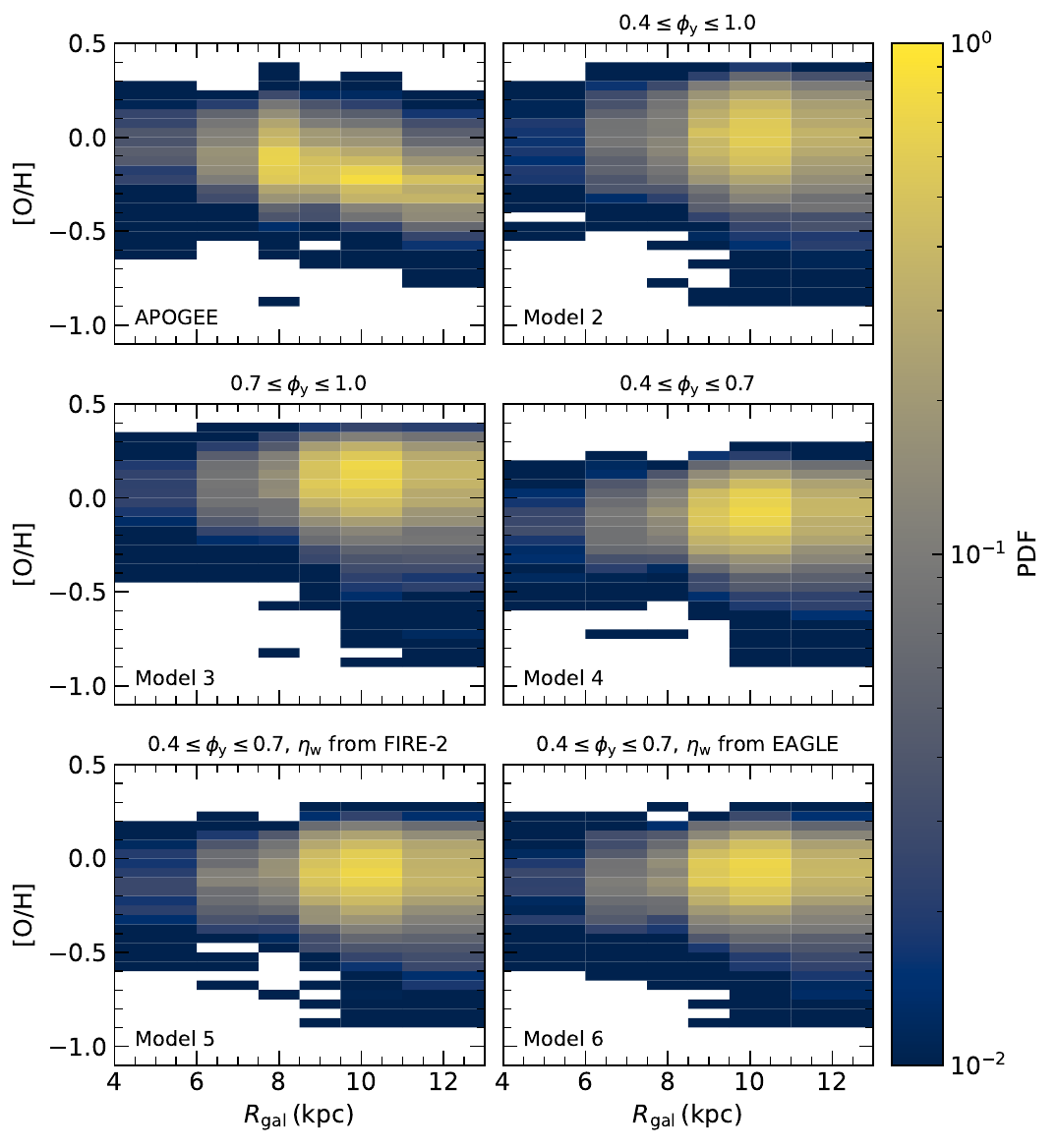}
\includegraphics[width=\columnwidth]{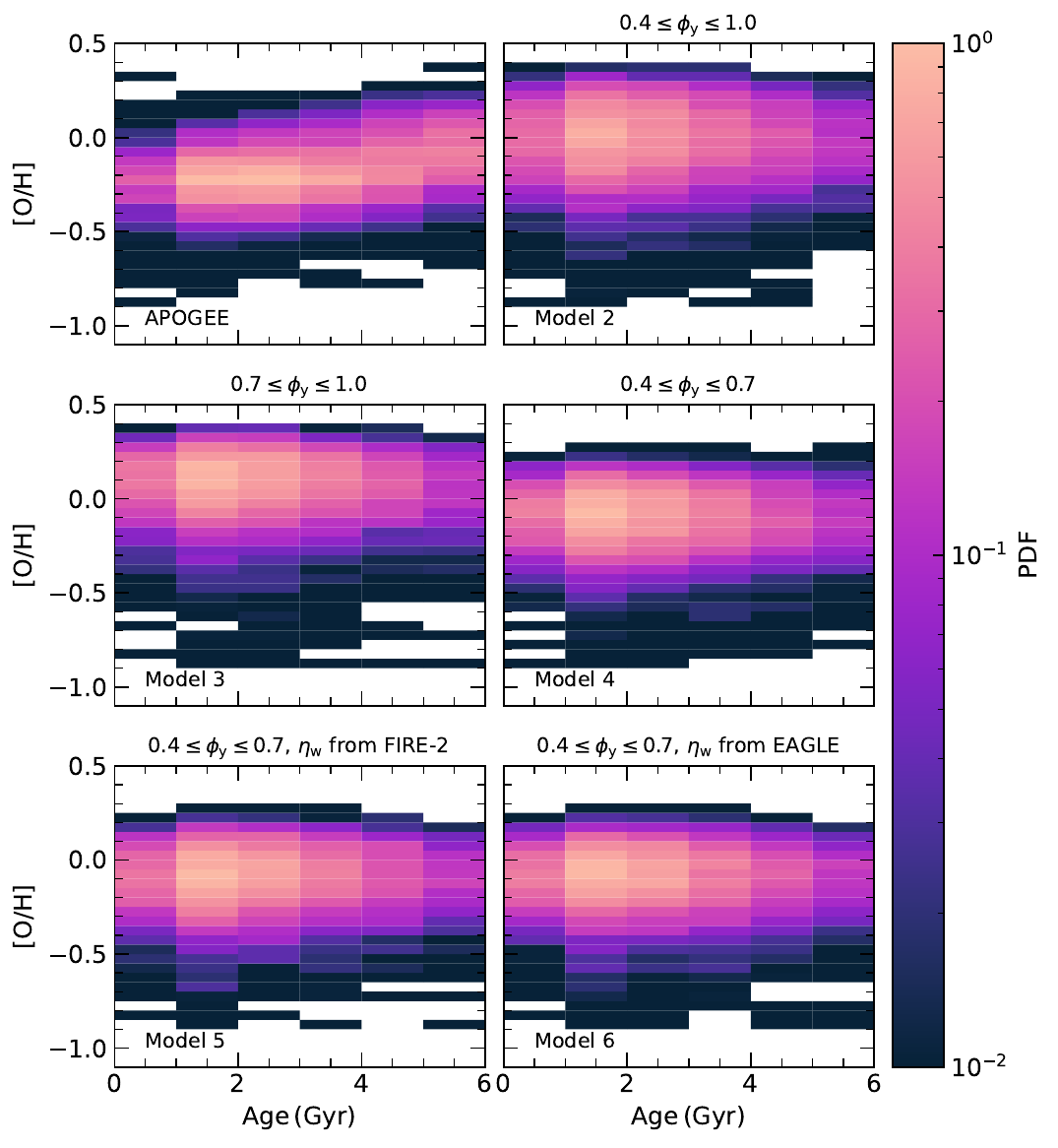}
\caption{Radius -- metallicity and age -- metallicity distributions for the sample of red clump stars from APOGEE DR14 used in this work (left panels), compared against the combination of the \protect\cite{2018ApJ...865...96F} stellar disc model and the fiducial ISM chemical evolution model from \protect\cite{2023arXiv230315853S}. The fiducial values of all ISM parameters are listed in \autoref{tab:taball}.}
\label{fig:fig4}
\end{figure*}

Following \cite{2018ApJ...865...96F}, we bin the APOGEE data in six bins between $4-13\,\rm{kpc}$, representing the present-day galactocentric radius of our sample of stars. Similarly, we bin the data in seven age bins spaced $1\,\rm{Gyr}$ apart. As we show in \aref{s:app_bins}, our results are robust against the number of bins used in either dimension. The left panels of \autoref{fig:fig3} show the APOGEE radius -- metallicity and age -- metallicity distributions. The corresponding right panels in \autoref{fig:fig3} show the same distributions from the fiducial model. We see from these panels that the model distributions are wider and more diffuse than the data distributions. The fiducial model produces more stars at low and high [O/H] at all ages. The overall metric value for the fiducial model is $\rm{\mathbb{M}} = 9.95$, with an uncertainty $\sim$ 0.3 that results from a combination of error on the ages and [O/H] as well as random error due to finite LHS sampling (\textit{i.e.,} $\mathbb{M}$ changes by approximately 0.3 when the number of LHS samples is either increased or decreased from 10000, or if a different set of 10000 randomly chosen grid points is used). 

It is not useful to gauge the performance of the fiducial model simply based on the metric value since we do not fit the model parameters. However, we can judge the \textit{relative} performance of different models by comparing their overall metric values. In fact, we will see below that some variations in the fiducial model produce a much better match with the data, whereas others perform poorly. In particular, we find four variations of the fiducial model to produce the lowest value on our metric (within the uncertainty) -- we refer to these as the best models. Below, we club and analyze results of each model we create based on the ISM physics they impact. 

\subsection{Differential metal enrichment of outflows}
\label{s:vary_phiy}
The first variations we introduce relate to the differential enrichment of galactic outflows, and are encoded in $\phi_{\rm{y}}$. The handful of nearby galaxies where outflow metallicity has been directly measured \citep{2018MNRAS.481.1690C} suggest that $\phi_{\rm{y}}$ increases with stellar mass \citep{2021MNRAS.504...53S}, implying that low mass galaxies preferentially lose metals via galactic outflows. 

$\phi_{\rm{y}}$ plays a central role in the ISM model, and the impact of variations in other parameters is limited as compared to $\phi_{\rm{y}}$. We thus opt to study different regimes of $\phi_{\rm{y}}$ rather than handpicking a few select values or scaling it up/down by some factor (as we do for the rest of the parameters). Models 2, 3, and 4 limit the range over which we vary $\phi_{\rm{y}}$ in the fiducial model. The limits we introduce ($\phi_{\rm{y}} = \{0.4, 0.7, 1.0\}$) are rather arbitrary, but capture distinct regimes of metal enrichment of outflows. We have checked that models with $\phi_{\rm{y}} \lesssim 0.4$ produce significantly more metal-poor stars than observed, so we do not discuss them in this work.

We present the resulting model distributions in \autoref{fig:fig4} and report the metric values in \autoref{tab:tab2}. We find that model 2 ($0.4 \leq \phi_{\rm{y}} \leq 1.0 $) and model 3 ($0.7 \leq \phi_{\rm{y}} \leq 1.0 $) perform slightly better and worse than the fiducial model, respectively. Interestingly, model 4 ($0.4 \leq \phi_{\rm{y}} \leq 0.7 $) performs much better as compared to models $1-3$, with a metric score roughly half of these models. This is also visible in the distributions for model 4, as we read off from \autoref{fig:fig4}. The effect of $\phi_{\rm{y}} < 1$ is that outflows remove a considerable fractions of metals from the disc, thereby making it difficult to produce stars with very high [O/H], which aligns with the observed distributions. We have checked that the best models prefer $0.4 \leq \phi_{\rm{y}} \leq 0.7 $ regardless of the value of (or variations in) other parameters. 

\subsection{Outflow mass loading factors}
\label{s:vary_etaw}
Based on the results from models 1 -- 4, we set a new baseline for creating more models, where we limit $0.4 \leq \phi_{\rm{y}} \leq 0.7 $. We have confirmed that models with all other variations that follow perform better when $\phi_{\rm{y}}$ is restricted to this range as compared to $\phi_{\rm{y}} < 0.4 $ or $\phi_{\rm{y}} > 0.7 $. In model 5, we set the outflow mass-loading factor, $\eta_{\rm{w}}\,(z)$, following results from the FIRE-2 simulations \citep{2021MNRAS.508.2979P}. Similarly, for model 6, we scale $\eta_{\rm{w}}\,(z)$ with stellar mass following the EAGLE cosmological simulations \citep{2020MNRAS.494.3971M}. For reference, at $z = 0$, $\eta_{\rm{w}} \approx 0.3$ from FIRE-2 and $\eta_{\rm{w}} \approx 1.0$ from EAGLE for a galaxy with $M_{\star} \approx 4\times10^{10}\,\rm{M_{\odot}}$ \citep[][figure 6]{2023arXiv230315853S}.

\autoref{fig:fig4} presents the resulting model distributions for model 5. It is difficult to ascertain whether models 5 and 6 perform better than model 4 by visual inspection. The value of the combined metric for models 4, 5 and 6 is 4.11, 4.83, and 4.67, respectively. Given the typical uncertainty of 0.3 associated with these values, we conclude that model 4 (with $\eta_{\rm{w}}$ from the analytical model of \citealt{2017MNRAS.465.1682H}) performs better than the others. However, the differences in the value of $\mathbb{M}$ between these three models is not as large as models 1 -- 4 where we varied $\phi_{\rm{y}}$ keeping $\eta_{\rm{w}}$ fixed. This finding is surprising because the evolution of $\eta_{\rm{w}}$ for the range in $M_{\star}$ we use in this work is different by a factor $2-10$ between the three scalings \citep[figure 6]{2023arXiv230315853S}. The relative insensitivity of the model distributions to $\eta_{\rm{w}}$ as compared to $\phi_{\rm{y}}$ implies that the mass loading of the outflows plays a less critical role in the models as compared to the metal mass loading. For subsequent models, we continue to use $\eta_{\rm{w}}\,(z)$ from the analytical model of \citeauthor{2017MNRAS.465.1682H}, because this model also gives $\eta_{\rm{w}} (z=0)$ and $\eta_{\rm{w}}(z)$ in good agreement with direct observations and previous modeling of the Milky Way \citep{2019ApJ...884...53F,2019ApJ...876...21P}.

\subsection{Gas accretion}
\label{s:vary_accretion}
Next, we vary two parameters that control the accretion of gas onto the disc: the halo concentration index $c$ and the baryonic accretion efficiency $\epsilon_{\rm{in}}$. Models 7 and 8 scale $c$ down and up by $2\times$, respectively, to examine if tuning this parameter provides a better match against the data. We show the resulting radius -- metallicity and age -- metallicity distributions in \autoref{fig:allothers} in \aref{s:app_MDFs}. The value of $\mathbb{M}$ is higher than the fiducial model in both cases, indicating a poorer reflection of the data. In particular, we find that the model age -- metallicity distribution is flatter than that observed when the concentration index is changed. In particular, model 8 produces more metal-rich stars at younger ages than required, which is reasonable because the dimensionless ratio $\mathcal{S}$ increases non-linearly with increasing $c$ via its influence on the rotational velocity $v_{\phi}$.  

Similarly, there is no guarantee that accretion in the Milky Way follows the fits used in the ISM model; in fact, the fit for $\epsilon_{\rm{in}}$ provided by \cite{2011MNRAS.417.2982F} that we use is only valid at $z \geq 2$, and $\epsilon_{\rm{in}}$ remains unconstrained at lower $z$ at least for the Milky Way. To test for variations in $\epsilon_{\rm{in}}$, models 9 and 10 scale it down and up by $2\times$, respectively. Both these models perform poorly on our metric; model 9 produces a lot more old and metal-rich stars whereas model 10 produces a substantial population of stars at low metallicities in the inner regions of the disc. These trends are expected because $\mathcal{A} \propto \epsilon_{\rm{in}}$, and decreasing $\mathcal{A}$ leads to higher [O/H] (see \autoref{s:models_ismmetallicity}). We thus conclude that variations in $c$ and $\epsilon_{\rm{in}}$ are not required to better reproduce the data.

\subsection{Gas fraction}
\label{s:vary_gasfraction}
Next, we examine parameters that determine the gas fraction in the disc, $f_{\rm{gQ}}$ and $f_{\rm{sf}}$. In model 11, we reduce $f_{\rm{gQ}}$ by $2\times$ whereas in model 12, we increase it by $1.5\times$.\footnote{We do not increase $f_{\rm{gQ}}$ by $2\times$ because this would give $f_{\rm{gQ}} > 0.95$, essentially implying that stars only contribute at less than 5 per cent level to the disc midplane pressure.} The resulting value of $\mathbb{M}$ for both models is greater than 10, indicating a much poorer fit to the data. Since both of these fractional parameters are directly proportional to the dimensionless ratio $\mathcal{S}$, all the stars in model 11 have [O/H] $< 0$, whereas model 12 produces a lot more metal-rich stars. This is perhaps not surprising given that the fiducial value of $f_{\rm{gQ}}$ we use is derived from the gas and stellar surface density of the Solar Neighbourhood \citep{2015ApJ...814...13M}. This mismatch therefore indirectly verifies the synergy between APOGEE data and gas-phase observations.

Instead of scaling $f_{\rm{sf}}$ by a constant factor, we produce a new model (model 13) where $f_{\rm{sf}}(z)$ is set by the analytic model of \citet{2009ApJ...693..216K} and \citet{2013MNRAS.436.2747K} that describes the fraction of star-forming gas as a function of local ISM properties. Model 13 performs the worst among all the models we investigate in this work. It gives a value of $f_{\rm{sf}} < 0.1$, and like model 11, does not produce any stars at [O/H] = 0 across the parameter space in radius and age. This is again due to a $f_{\rm{sf}} < 0.1$ yielding an unreasonably low $\mathcal{S}$ that brings down [O/H]. We thus conclude that the fiducial choice for $f_{\rm{sf}}$ is adequate to describe the fraction of star-forming gas in the Milky Way.


\subsection{Disc scale height}
\label{s:vary_scaleheight}
Given that the ISM model invokes vertical hydrostatic equilibrium, we can also test for the impact of the scale height of the disc on the model distributions. The scale height is given by the combination of the gas surface density and velocity dispersion \citep[][equation 24]{2018MNRAS.477.2716K}. For this purpose, we turn to the parameters $\eta$ and $\phi_{\rm{mp}}$ since the model assumes that the turbulence crossing time is of the order of the gas scale height. In model 14, we increase $\eta$ to 3, such that the energy injected in the ISM is radiated away in roughly two crossing times. In model 15 (16), we decrease (increase) the value of $\phi_{\rm{mp}}$ to 1 (2). $\phi_{\rm{mp}} = 2$ means that the cosmic ray pressure in the disc midplane is equivalent to magnetic pressure, whereas $\phi_{\rm{mp}} = 1$ implies that the thermal pressure is non-negligible.

Models 14 and 15 perform slightly worse than model 4 (the fiducial model with $0.4 \leq \phi_{\rm{y}} \leq 0.7 $) on our metric, indicating that (1.) our assumption that turbulence is dissipated in a single crossing time is reasonable, and (2.) thermal pressure at the midplane is negligible in the Milky Way, as is expected for massive spiral galaxies. The scale heights we obtain for the disc vary from $0.1\,\rm{kpc}$ in the inner regions to $0.4\,\rm{kpc}$ in the outer regions.

\begin{figure*}
\includegraphics[width=\columnwidth]{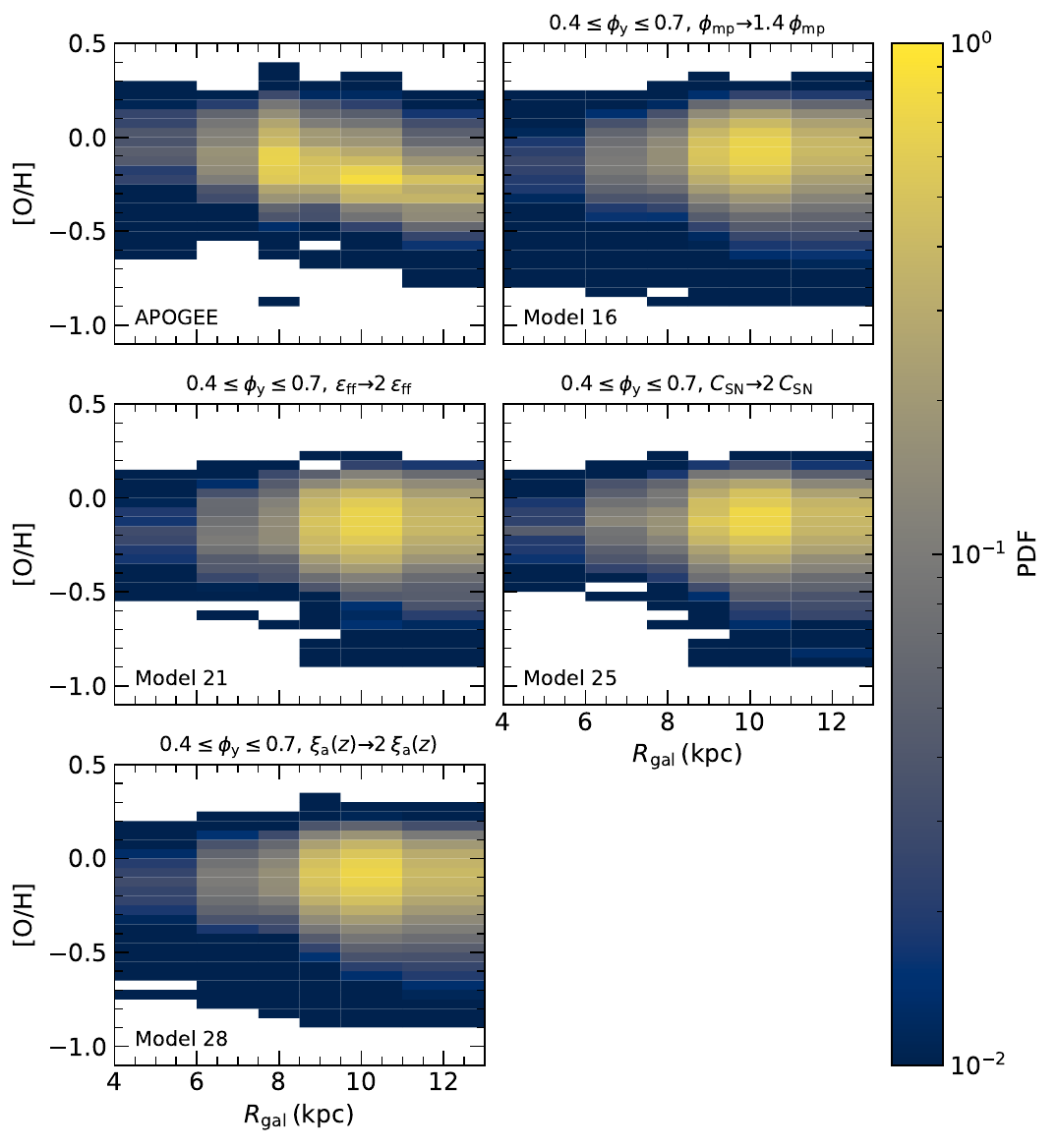}
\includegraphics[width=\columnwidth]{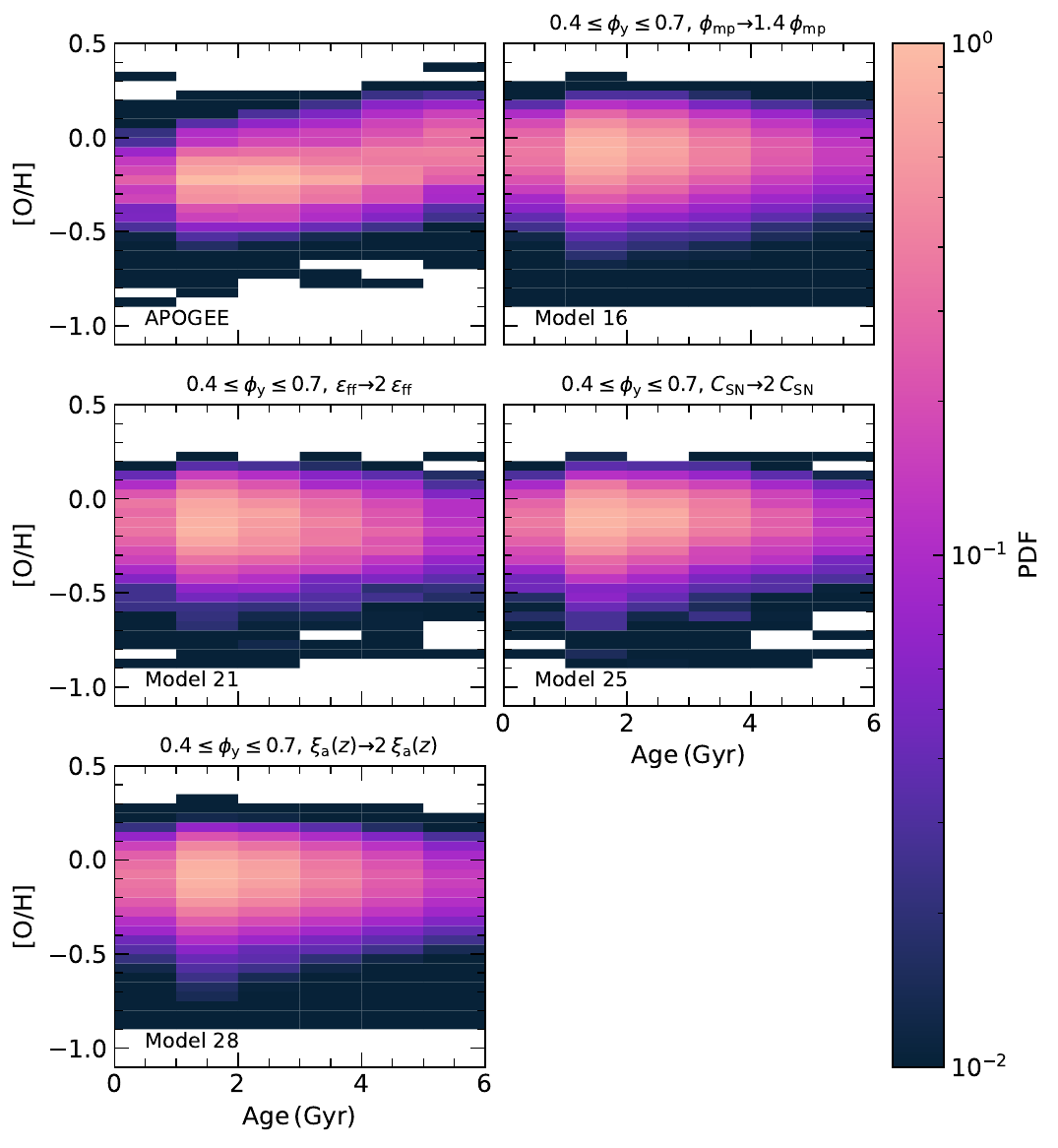}
\caption{Same as \autoref{fig:fig4}, but for the models that perform the best on the metric $\mathbb{M}$ used in this work. The differences between these models and the fiducial model are listed in \autoref{tab:tab2}.}
\label{fig:fig5}
\end{figure*}

We show the distributions for model 16 in \autoref{fig:fig5}. The distributions from model 16 are slightly broader as compared to the data, especially for older stars as well as stars located at large radii. Nevertheless, model 16 performs as well as model 4 within the uncertainty. While the dynamic range of $\phi_{\rm{mp}}$ is rather limited, the comparable performances of models 4 and 16 imply that cosmic ray pressure plays a non-negligible role in setting the ISM pressure in the disc. Model 16 is one of the four best models we find in our analysis.  

\subsection{Disc self gravity}
\label{s:vary_selfgravity}
The self-gravity of the disc determines its stability against gravitational collapse, and is thus an important property that dictates the onset of star formation and metal production. The two model parameters that set the self-gravity of the gas are $Q_{\rm{min}}$ and $\phi_{\rm{Q}}$. In the ISM model, we assume that galaxies can self-regulate $Q$ to maintain a balance between star formation, gas accretion, outflows, and transport. Simulations and observations typically find that $Q_{\rm{min}}$ varies between $1-3$ on orbital timescales \citep[e.g.,][]{2008AJ....136.2782L,2014ApJ...785...75G,2016MNRAS.456.2052I}, so we introduce two new models where $Q_{\rm{min}} = 1$ (model 17) and 3 (model 18). Similarly, we increase $\phi_{\rm{Q}}$ to 4 in model 19 to test if a higher Toomre $Q$ for gas results in a better match with the data.

From \autoref{fig:allothers}, we find that model 17 shows an overabundance of metal-rich stars whereas model 18 shows the opposite. It is not straightforward to find out the reason for these trends because \textit{if} scaling $Q_{\rm{min}}$ shuts off transport, then $Q$ can take on any value equal or larger than $Q_{\rm{min}}$, and $\mathcal{S} \propto Q^{-2}$ whereas $\mathcal{P} \propto Q^{-2}_{\rm{min}}$. Nevertheless, $\mathbb{M} > 10$ for both models 17 and 18, implying that they can not explain the data. Model 19 performs better than models 17 and 18, however it lags behind model 4. This is expected since both $\mathcal{S}$ and $\mathcal{A} \propto \phi_{\rm{Q}}$ but $\mathcal{P} \propto \phi^2_{\rm{Q}}$. These experiments thus suggest that the fiducial values of $Q_{\rm{min}}$ and $\phi_{\rm{Q}}$ are adequate to reproduce the observed radius -- metallicity and age -- metallicity distributions.


\subsection{Star formation}
\label{s:vary_starformation}
To test for the effects of variations in our star formation prescription, we first vary the star formation efficiency per free-fall time by $2\times$. While observations find $\epsilon_{\rm{ff}} \approx 0.01$ across orders of magnitude in gas density and star formation rate \citep[e.g.,][]{2012ApJ...745...69K,2015ApJ...806L..36S,2018MNRAS.477.4380S,2019MNRAS.487.4305S}, more precise measurements in resolved clouds reveal an intrinsic scatter in $\epsilon_{\rm{ff}}$ by a factor of $2$ \citep{2022MNRAS.511.1431H}. We find that model 20, where $\epsilon_{\rm{ff}}$ is reduced by $2\times$, performs somewhat poorly as compared to model 4, as it contains a higher fraction of somewhat metal-poor stars than the data. 

However, model 21 where we increase $\epsilon_{\rm{ff}}$ by $2\times$ outperforms model 4, and is one of the four best models. We show the distributions for model 21 in \autoref{fig:fig5}. We see that while model 21 nicely captures the extent and shape of the observed radius -- metallicity distribution, it does not well reproduce the age -- metallicity distribution seen in the data. Nevertheless, the overall value of the metric $\mathbb{M}$ is smaller than that of model 4. This result highlights the possibility that stars formed with a slightly higher efficiency per free-fall time in the Galactic disc, which we discuss below in \autoref{s:discussion}. 

As we mention in \autoref{s:models_overview}, our ISM model includes two modes of star formation -- one where it exclusively occurs in GMCs that are decoupled from the disc, and another where the entire volume of the disc can undergo star formation (the so-called Toomre regime of star formation, commonly seen in ULIRGs and starbursts). The transition from one regime to the other is determined by the maximum gas depletion timescale, $t_{\rm{sf,max}}$ in the model -- for large values of $t_{\rm{sf,max}}$, star formation occurs in a continuous medium and is dictated by the free-fall time of the gas. On the other hand, if the gas depletion timescale is shorter than the free-fall timescale, stars forms in clouds which are collapsing at the same as the gas is being depleted. These scenarios are depicted by models 22 and 23, where we scale $t_{\rm{sf,max}}$ down and up by a factor of 2, respectively. A lower value of $t_{\rm{sf,max}}$ will result in a higher star formation rate, thus, model 22 produces a lot more metal-rich stars whereas model 23 contains no stars with [O/H] > 0. Unsurprisingly, both these models have $\mathbb{M} > 10$, and fail to reproduce the data. This failure is also consistent with observational constraints on $t_{\rm{sf,max}}$ that find it to be $\sim 2\,\rm{Gyr}$ for local galaxies \citep{2008AJ....136.2782L,2008AJ....136.2846B,2020ApJ...901L...8S}.

Finally, in model 24, we switch off the GMC regime, thus mimicking all other chemical evolution models that only consider the Toomre regime (or an even simpler prescription for star formation). The resulting value of $\mathbb{M} > 40$, which is not surprising, as \cite{2018MNRAS.477.2716K} point out the relevance of the GMC regime of star formation for Milky Way-like galaxies. This is critical to take into account because not all the available gas in the Milky Way is forming stars, and most star formation is occurring exclusively in GMCs. 

\subsection{ISM turbulence}
\label{s:vary_turbulence}
One of the key strengths of our ISM model lies in the connection between gas dynamics and metallicity distribution, which, despite longstanding arguments \citep{1980FCPh....5..287T}, has not been explored except in a handful of cosmological simulations \citep[e.g.,][]{2017MNRAS.466.4780M,2021MNRAS.506.3024H}. In our ISM model, turbulence can be injected into the ISM by supernova feedback, gas accretion, or radial gas flows. The resulting gas velocity dispersion influences all the three dimensionless ratios: $\mathcal{P} \appropto \sigma_{\rm{g}}$, $\mathcal{S} \propto \sigma^{-2}_{\rm{g}}$, and $\mathcal{A} \propto \sigma^{-3}_{\rm{g}}$.


In models 25 and 26, we take supernova clustering into account by simply increasing $C_{\rm{SN}}$ to $2$ and $5$, respectively. We notice that more intense feedback due to clustering yields a smaller value of $\mathbb{M}$ as compared to model 4. We show the distributions for models 25 and 26 in \autoref{fig:fig5}. While both models 25 and model 26 can reasonably reproduce the data, we will show later in \autoref{s:discussion} that an excessively large $C_{\rm{SN}}$ violates observational constraints on the gas velocity dispersion. Hence, our conclusion is that while feedback from clustered supernovae offers a more favorable fit to the APOGEE data, very aggressive feedback is likely implausible as it leads to an overestimation of the gas velocity dispersion.

Similarly, we also investigate if varying the turbulent injection efficiency due to accretion ($\xi_{\rm{a}}$) has an impact on our results. Models 27 and 28 use a value of $\xi_{\rm{a}}$ that is scaled down and up by $2\times$ as compared to the fiducial choice, respectively. While model 27 performs slightly worse than model 4, model 28 performs much better, and is the last of the four best models we find. From \autoref{fig:fig5}, we find that the distributions from model 28 are akin to those from model 16 where we took into account the contribution from cosmic ray pressure at the disc midplane. Overall, models 25, 26, and 28 favor a slightly higher level of turbulence ($\sigma_{\rm{g}}$). 

\subsection{Radial gas flows}
\label{s:vary_gasflows}
Our ISM model includes a self-consistent treatment of radial gas flows.\footnote{Note that our definition of radial gas flows is fundamentally different from similar models developed for the Milky Way \citep[e.g.,][]{2023MNRAS.523.3791C}. These models solve for chemical evolution in distinct radial zones while treating the radial flow of gas from the outskirts to the center post accretion. We incorporate radial flows that would be necessary to achieve equilibrium \textit{after} the accreted gas has already been transported down the potential well of the galaxy.} In equilibrium, radial flows turns off if turbulence present in the ISM can be sustained by star formation feedback and gas accretion. We find that some models show radial gas flow and metal advection at some epochs, while it is switched off in other models. Further, radial gas flow is also sensitive to the size of the disc; since we sample over the disc size, we find that gas flows are present for some disc sizes but not for others for the \textit{same} model. Interestingly, we find that radial gas flows are inactive for all $R$ for the four best models we find in this work. This is not surprising, given that radial gas flows are not always observed in local spiral galaxies \citep{2021ApJ...923..220D}. From the point of view of equilibrium models, this is because turbulence can be maintained by accretion and supernova feedback in these galaxies \citep{2018MNRAS.477.2716K,2020A&A...641A..70B}. We discuss this further below in \autoref{s:discussion}.


\begin{figure}
\includegraphics[width=\columnwidth]{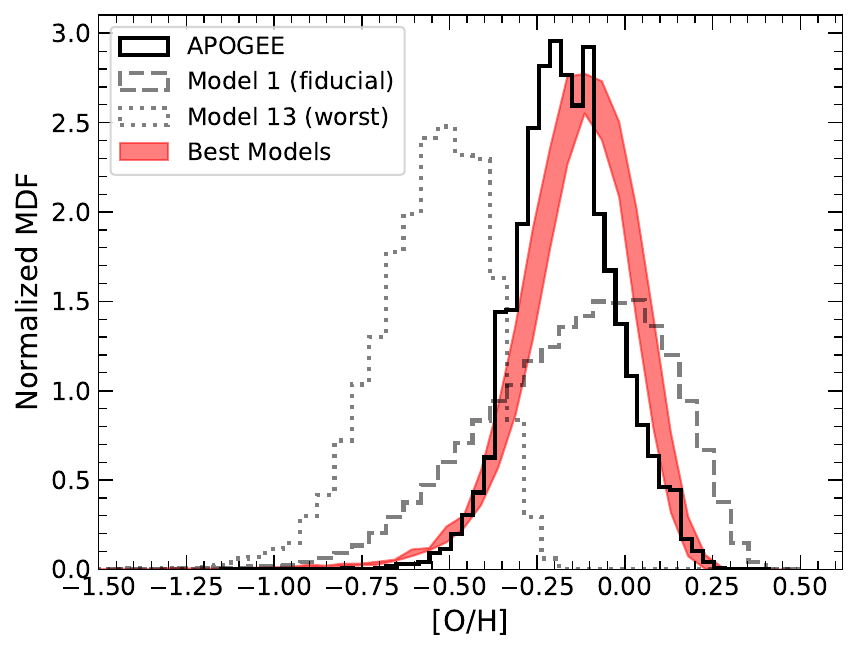}
\caption{Normalized metallicity distribution function (MDF) for the APOGEE data, plotted together with the normalized MDFs from the four best models (models 16, 21, 25, and 28 -- see \autoref{fig:fig5}). For comparison, also shown are the normalized MDFs for the fiducial model and the model with the worst score on the metric (model 13). A description of all the models is available in \autoref{tab:tab2}.}
\label{fig:figlast}
\end{figure}

\subsection{Summary of best models}
\label{s:vary_summary}
To summarize, we find four models that outperform the fiducial model. These are model numbers 16, 21, 25 and 28 (see \autoref{tab:tab2}); they have the least value of the metric $\mathbb{M}$. In model 16, we scaled $\phi_{\rm{mp}}$ to 2, thereby including the contribution of cosmic rays to the midplane pressure. In models 21 and 25, we increased the star formation efficiency and supernovae clustering, respectively. Finally, in model 28, we tuned up the level of turbulence injected into the ISM by gas accretion. All these variations provide interesting insights into the ISM history of the Milky Way, which we discuss in detail below. To aid in comparison, we show the normalized MDFs for these models with the APOGEE MDF and the MDF from the fiducial and the worst models in \autoref{fig:figlast}. It is immediately obvious that the fiducial model produces a broader MDF than the data, whereas the worst model (model 13) produces a much more metal-poor MDF than the data. The most important result of our analysis is that all the best models prefer $0.4 \leq \phi_{\rm{y}} \leq 0.7 $, and it seems indispensable to \textit{not} assume $\phi_{\rm{y}} = 1$ or not include it in chemical evolution models, which is the usual practice. 

As we read off from \autoref{fig:figlast}, the MDFs for the best models overlap the most with the data MDF, but there is clearly scope for further tuning the model parameters to better reproduce the observed MDF. Given our objective, we have only experimented with very simple variations in the ISM model parameters, and refrained from testing complex, multi-parameter scalings as a function of radius or age. Conversely, we do not expect to reproduce detailed features of the observed MDF. This exercise will be the subject of a forthcoming study.

\section{Discussion}
\label{s:discussion}
We have seen in \autoref{s:analysis} how varying the ISM parameters impacts the performance of our models against the APOGEE data. We first discuss the common features of the four best models we find in our analysis (summarized in \autoref{s:vary_summary}), and argue that they also satisfy constraints from ISM observations for the Milky Way, or, from local spiral galaxies that are expected to evolve in a similar manner to the Milky Way. This is an important sanity check because there is no guarantee that models that well reproduce the stellar distributions do not violate gas-phase observations. Then, we discuss predictions from the best models for parameters that have little to no observational constraints from external galaxies, and provide the first insights into them using Milky Way data.

\subsection{Consistency with gas-phase observations}
\label{s:consistency}

\begin{figure}
\includegraphics[width=\columnwidth]{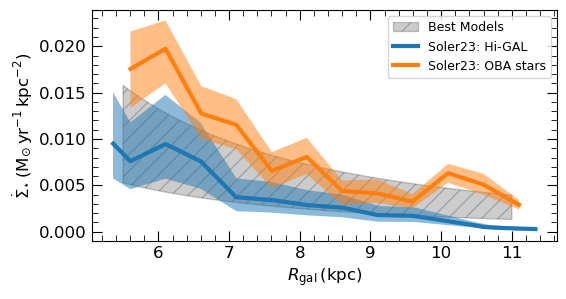}
\caption{Radial profiles of the star formation rate surface density in the Galaxy, adopted from \citep{2023A&A...678A..95S}. The blue curve corresponds to $\dot \Sigma_{\star}$ measured from protostellar clumps in the Hi-GAL survey \citep{2021MNRAS.504.2742E,2022ApJ...941..162E}. The range curve denotes $\dot \Sigma_{\star}$ inferred from massive young stars within $6\,\rm{kpc}$ of the Sun \citep{2023A&A...669A..10Z}. In both cases, the shaded regions denote the $16^{\rm{th}}$ and $84^{\rm{th}}$ percentiles. Overplotted in grey is the range of $\dot \Sigma_{\star}$ in the best models.}
\label{fig:fignew8}
\end{figure}

\textbf{Mass flow rates} First, we look at the mass flow rates as they set the overall mass balance in the ISM model. The present day star formation rates for the best models is $1-3\,\rm{M_{\odot}\,yr^{-1}}$, in excellent agreement with existing multi-wavelength observations \citep[table 1, and references therein]{2022ApJ...941..162E}. In addition to the global SFR, we also compare the radial profile of the SFR surface density predicted by the best models with estimates from two different observations in \autoref{fig:fignew8}. The observational estimates, presented in \citet{2023A&A...678A..95S}, are derived from protostellar clumps emitting in the infrared in the Hi-GAL survey \citep{2021MNRAS.504.2742E,2022ApJ...941..162E} that covered a 2 degree strip around the Galactic midplane, and from massive young stars within a $6\,\rm{kpc} \times 6\,\rm{kpc}$ area centered on the Sun \citep{2023A&A...669A..10Z}. Our predictions better agree with $\dot \Sigma_{\star}$ from the Hi-GAL survey, which is lower than $\dot \Sigma_{\star}$ from \citet{2023A&A...669A..10Z} by a factor of two. Given the systematic offset between the two datasets (see \citealt{2023A&A...678A..95S} for a discussion on possible reasons for the offset) and the qualitative nature of this work,  the agreement between the modelled and observed $\dot \Sigma_{\star}$ profiles is encouraging, and constitutes an important step in validating the models. We also obtain gas accretion rates of the same magnitude in these models. It is non trivial to compare the model accretion rates with direct observations because the latter decompose the accreting gas into different thermal phases, and at different locations throughout the Galactic halo. Nonetheless, these observations estimate gas accretion rate for the Galaxy to be $0.2-3\,\rm{M_{\odot}\,yr^{-1}}$ \citep[e.g.,][]{2012ARA&A..50..491P,2012ApJ...750..165R,2017ASSL..430...15R,2014ApJ...787..147F,2019ApJ...884...53F}, consistent with our models. As we mention in \autoref{s:vary_etaw}, we also obtain mass loading factors in agreement with estimations for the Milky Way \citep{2019ApJ...884...53F,2019ApJ...876...21P}, although these measurements suffer from the same problem as the accretion rates. Given the depth of available data, we can confirm that the large-scale mass flow rates for the best models do not violate existing observational constraints. 

\textbf{GMC regime of star formation} It is well known that star formation largely occurs in GMCs in the Milky Way \citep{1993prpl.conf..125B,2008AJ....136.2782L}, and we take this into account in the ISM model via the GMC regime of star formation. The importance of the GMC regime is also reflected in all the best models, and in the the large metric value for model 24 where we exclude it. Moreover, the data prefer the fiducial value of the timescale for star formation in the GMC regime, $t_{\rm{sf,max}}$, which is adopted from the gas depletion timescale in the Milky Way and nearby local spirals \citep{2008AJ....136.2782L,2020ApJ...901L...8S,2022MNRAS.516.3006K}. Similarly, it is encouraging that models that use the fiducial value of the effective gas fraction in the disc, $f_{\rm{gQ}}$, perform better than models where it is varied, as the fiducial value is directly obtained from measurements within the Solar neighborhood \citep{2015ApJ...814...13M}.

\textbf{Star formation efficiency} One of the models that best reproduces the data is model 21, where we set $\epsilon_{\rm{ff}} = 0.03$, a factor of 2 higher than the typical value \citep{2012ApJ...745...69K}. While a factor of 2 variation in $\epsilon_{\rm{ff}}$ is generally not considered significant as it is within observational uncertainties \citep[e.g.,][]{2022MNRAS.511.1431H}, very precise measurements of the gas surface density and star formation rate surface density in nearby Milky Way molecular clouds indeed find $\epsilon_{\rm{ff}} \approx 0.026$ \citep{2021ApJ...912L..19P}. This consistency between our models that best reproduce the APOGEE data and direct observations of nearby star-forming molecular clouds strongly indicates that stars have formed more efficiently (in a free-fall time) in the Milky Way, at least in its recent history.

\textbf{Hydrostatic equilibrium} We also see that model 16 (with a lower value of $\phi_{\rm{mp}}$) outperforms other models, although varying $\phi_{\rm{mp}}$ has a limited impact on all the metrics. The higher value of $\phi_{\rm{mp}}$ is suggestive of the non-negligible role of cosmic ray pressure (as compared to magnetic pressure) in setting hydrostatic equilibrium in the Galactic disc, and is supported by both Galactic observations and theory \citep{1990ApJ...365..544B,2021MNRAS.502.1312C}. Furthermore, the scale heights we obtain for the disc (that depends on the gas surface density and velocity dispersion -- see equation 24 of \citealt{2018MNRAS.477.2716K}) vary from $0.1\,\rm{kpc}$ in the inner regions to $0.4\,\rm{kpc}$ in the outer regions, in good agreement with those inferred by \citet[figure 1]{2019A&A...632A.127B} for the Galaxy using classical Cepheids (\citealt{2019NatAs...3..320C}; see also, \citealt{2022MNRAS.513.4130L} and \citealt{2024arXiv240105023C}).

Overall, we find that our best models are consistent with a large set of gas-phase observations of the Milky Way, and can simultaneously reproduce the observed 2D distributions of the metallicity of red clump stars as a function of Galactocentric distance and age. To highlight the predictive power of the models, we next investigate model parameters that are also well constrained by the APOGEE data but have no direct, gas-phase observational constraints as of yet.

\begin{figure*}
\includegraphics[width=\textwidth]{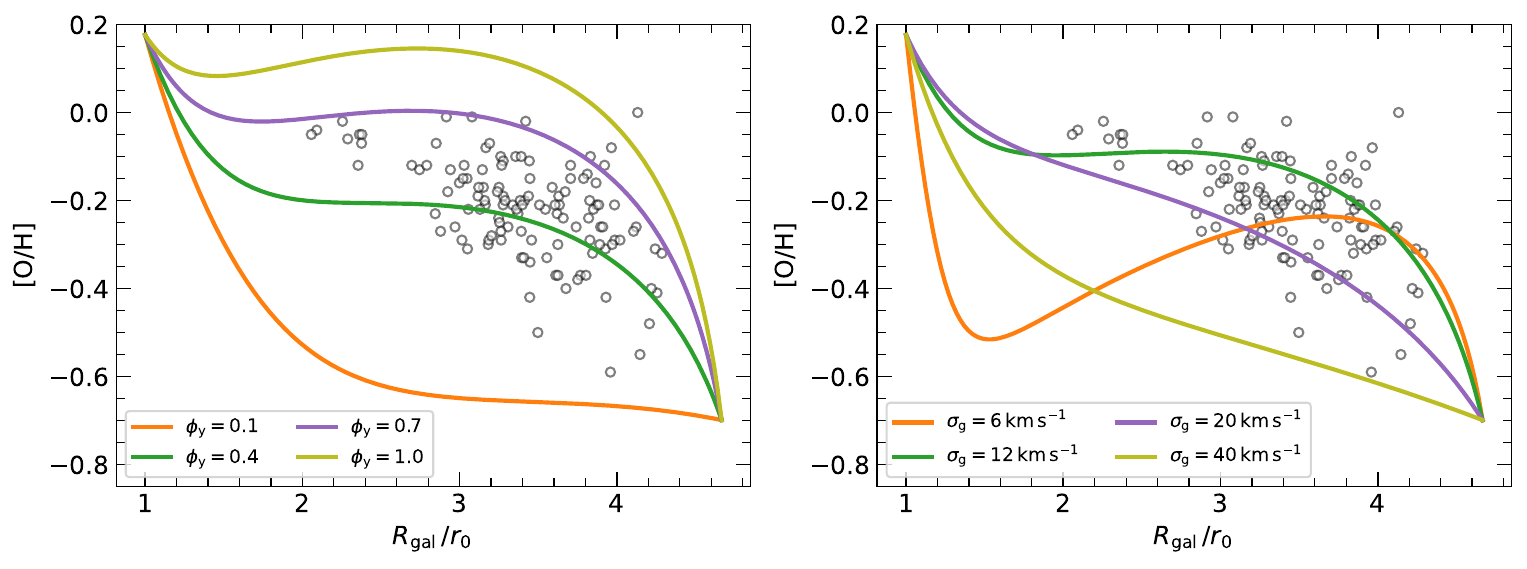}
\caption{\textit{Left panel:} Oxygen abundance as a function of normalized Galactocentric radius ($r_0 = 3\,\rm{kpc}$). Grey markers denote the subset of the youngest, low$-\alpha$ red clump stars (ages $< 0.5\,\rm{Gyr}$) in APOGEE DR14 data. Solid curves denote the metallicity profile predicted by the fiducial model with varying levels of metal enrichment of Galactic outflows, encoded in the fractional parameter $\phi_{\rm{y}}$. $\phi_{\rm{y}} = 1$ means that the metallicity of the outflows is equal to the ISM metallicity, whereas $\phi_{\rm{y}} \ll 1$ means most of the newly produced type II metals are ejected via outflows. Most of the data prefer an intermediate $\phi_{\rm{y}}$, as we discuss in the text. \textit{Right panel:} Same data as the left panel, but the models shown are for varying levels of ISM turbulence (encoded in the gas velocity dispersion $\sigma_{\rm{g}}$) for fixed $\phi_{\rm{y}} = 0.5$. The data prefer intermediate values of $\sigma_{\rm{g}}$.}
\label{fig:fig8}
\end{figure*}

\subsection{Insights into observationally-unconstrained ISM parameters}
\label{s:predictions}
Three parameters in the ISM model have no direct constraints from observations of the Milky Way or its analogues. These are -- 1.) preferential metal enrichment of galactic outflows ($\phi_{\rm{y}}$), 2.) supernovae clustering ($C_{\rm{SN}}$), and 3.) accretion-driven turbulence efficiency ($\xi_{\rm{a}}$). $\phi_{\rm{y}}$ dictates the metallicity of galactic outflows, while the other two enhance turbulence in the disc, thereby increasing the gas velocity dispersion.

We concluded in \autoref{s:analysis} that all the best models require $0.4 \leq \phi_{\rm{y}} \leq 0.7$. One might worry that any conclusions about $\phi_{\rm{y}}$ are potentially sensitive to the assumed IMF-averaged yield $y$, which is uncertain by a factor $\sim 3$ for oxygen. This uncertainty arises from our incomplete knowledge of stellar nucleosynthesis \citep{2020ApJ...900..179K,2022A&ARv..30....7R}, which massive stars go supernovae \citep{2003ApJ...591..288H,2016ApJ...821...38S}, effects of pre-supernova enrichment \citep{2021A&A...656A..58L}, etc. If the IMF systematically varies as a function of metallicity (as is expected from ISM thermodynamics -- \citealt{2001MNRAS.322..231K,2022MNRAS.509.1959S}), this will also affect $y$. We have used a constant $y/\rm{Z_{\odot}} \approx 2$ in this work. If we vary this ratio between $1-3$ \citep[e.g.,][figure 1]{2023arXiv230609133W} and re-consider variations in the fiducial model, we still find $0.4 \leq \phi_{\rm{y}} \leq 0.7$ to best describe the data. This implies that in order to reproduce the observed distributions, the ISM model requires the existence of metal-rich outflows (relative to the local ISM) from the disc over the past 6 Gyr. Based on the mass-loading factor we obtain from \cite{2017MNRAS.465.1682H}, this range of $\phi_{\rm{y}}$ corresponds to outflow metallicities $1.2-1.8\times$ local ISM metallicity when integrated across the disc. 

It is well known that superbubbles seen in nearby spiral galaxies are a result of clustered supernovae explosions \citep[e.g.,][]{1988ApJ...324..776M,2017MNRAS.465.1720Y,2019MNRAS.490.1961E,2023A&A...676A..67W}. It has been long suspected that superwinds created due to clustered supernova feedback might be driving metal-enriched outflows from the Solar neighbourhood (e.g., \citealt{1992A&A...266..190M}; see also, \citealt{2018MNRAS.481.3325F}) and impacting subsequent star formation \citep{2013ApJ...775..110H,2022Natur.601..334Z}. Our finding that $\phi_{\rm{y}} < 1$ for the Milky Way is also supported by isolated galaxy formation simulations with resolution high enough to resolve metal mixing at the ISM -- CGM interface \citep{2023arXiv230907955V}. These simulations estimate $\phi_{\rm{y}} < 1$ even for Solar Neighborhood conditions. However, considerable uncertainties remain in directly measuring the metallicity of the outflowing gas \citep[e.g.,][]{2017MNRAS.469.4831C}, not to mention the added complication due to possible re-accretion of metal-rich gas via Galactic fountains \citep[e.g.,][]{2016MNRAS.456.2140M,2016ApJ...816L..11F,2017MNRAS.470.4698A,2020A&A...642A.163S}. Nonetheless, our results point that even for a massive galaxy like the Milky Way, the commonly used assumption that the outflow metallicity equals the ISM metallicity may not be correct, and should be revisited in Galactic chemical evolution models. If the Galaxy indeed preferentially expels gas rich in oxygen (more generally, type II elements), this also has implications for the evolution of O/Fe, C/O and N/O ratios as this loss is not taken into account in commonly accepted explanations for the CNO cycle or the $\alpha-$ knee \citep[see the review by][]{2022A&ARv..30....7R}.

We also infer from \autoref{s:analysis} that models that either include feedback from clustered supernovae or a higher turbulence injection efficiency from gas accretion better reproduce the APOGEE distributions. Note that large values of both $C_{\rm{SN}}$ and $\xi_{\rm{a}}$ increase the velocity dispersion of the gas. The elevated velocity dispersion (averaged over the last 6 Gyr) is higher than that seen in a typical galaxy of similar mass but within the scatter present in the trend between stellar mass and gas velocity dispersion \citep{2020MNRAS.495.2265V}. Higher $\sigma_{\rm{g}}$ in turn increases $\mathcal{P}$ but decreases $\mathcal{S}$ and $\mathcal{A}$ in the metallicity equation (see \autoref{eq:P}). However, model 26 (with $C_{\rm{SN}} = 5$) increases $\sigma_{\rm{g}}$ to $50\,\rm{km\,s^{-1}}$ at $z = 0$, which is larger by a factor $\sim 4$ than that observed in the Galactic ISM \citep{2009ARA&A..47...27K,2020A&A...633A..14R}, and even larger than the stellar velocity dispersion of stars younger than 6 Gyr \citep{2019MNRAS.489..176M}. Therefore, even though model 26 has an $\mathbb{M}$ at par with the other best models, it severely violates observational constraints on both stellar and gas-phase velocity dispersions. Together with the fact that model 25 ($\mathbb{M} = 4.11 \pm 0.3$) performs similar to model 4 ($\mathbb{M} = 4.19 \pm 0.3$), we conclude that the models allow for, at most, some level of supernova clustering to reproduce the radius -- metallicity and age -- metallicity distributions, but do not require it. It is quite plausible that supernova clustering can simultaneously increase ISM turbulence and metal content of outflows (e.g., \citealt{2018ApJ...869...94E,2018MNRAS.481.3325F,2020ApJ...895...43S,2020ApJ...903L..34K}; Sharda et al. 2024, in prep.), both of which are preferred by models best describing the APOGEE data.

On the other hand, model 28 with $\xi_{\rm{a}} (z)$ twice the fiducial value yields $\sigma_{\rm{g}} (z=0)$ consistent with observations, while also reproducing the observed APOGEE distributions. $\sigma_{\rm{g}}$ in model 28 is higher by 20 per cent than the average trend (\citealt{2019ApJ...880...48U,2019ApJ...886..124W}, Wisnioski et al. in preparation) at the highest redshifts we consider. Our findings suggest that gas accretion could have driven a non-negligible amount of turbulence in the Galactic disc over the past 6 Gyr. This has not been investigated before because chemical evolution models typically do not take into account energy balance in the ISM in addition to mass and metal balance. However, there are no direct observational measurements of $\xi_{\rm{a}}$, and simulations have only explored it for dwarf galaxies \citep{2022arXiv220405344F}. Another implication of our results is on the age -- velocity dispersion relation (AVR) for the Milky Way disc, and whether stars were born \textit{hot} or were dynamically heated afterwards \citep[e.g.,][]{1951ApJ...114..385S,2002MNRAS.336..785S}. If the ISM in the Galaxy had higher than average velocity dispersion in the past, it would mean that the stars were born hotter and contribution from dynamical heating to their observed velocity dispersions is less prominent than expected \citep[e.g.,][]{2012MNRAS.420..913B,2013ApJ...773...43B,2017MNRAS.472.1879L,2019MNRAS.489..176M,2019ApJ...878...21T}.  

Lastly, we discuss the role of metal diffusion in our models. The diffusion coefficient in the models is proportional to $\sigma^2_{\rm{g}}$, and increases linearly with radius under the Toomre stability criterion (see section 2.2 of \citealt{2021MNRAS.502.5935S} for details). Intuitively, we can understand this as follows: high $\sigma_{\rm{g}}$ inevitably leads to more metal mixing due to turbulence or other instabilities \citep{2017MNRAS.466.4780M,2021MNRAS.506.1295S}, and the outer regions of galaxies have larger scale heights due to flaring \citep{2019A&A...632A.127B}, which makes it possible for diffusion to act on larger scales. Thus, diffusion becomes dominant (when any of the three dimensionless ratios $\mathcal{P}, \mathcal{S}$ and $\mathcal{A}$ are less than unity) in redistributing metals at earlier times and at larger radii.

While we explore ISM physics in a lot more detail as compared to other studies, our investigation solely focuses on the oxygen abundance. This trade-off prevents us from leveraging the abundances of various other elements in the APOGEE sample, which can provide valuable insights into different nucleosynthesis pathways and their impact on the ISM \citep[e.g.,][]{2001ApJ...554.1044C,2010A&A...522A..32R,2017A&A...605A..59R,2018MNRAS.475.2236K,2021MNRAS.507.5882S,2022ApJ...927..209T}.  



\subsection{Which aspects in APOGEE data constrain outflow metallicity and ISM turbulence?}
\label{s:whichaspects}
Given that our models find Galactic outflows to be metal enriched and gas velocity dispersion to be slightly elevated, it is worth asking which particular aspects of the APOGEE data we use are ultimately responsible for these conclusions.

To understand this, we revisit the metallicity equations (\autoref{eq:Z_GMC} and \autoref{eq:Z_Toomre}), and analyze the impact of the outflow metallicity parameter $\phi_{\rm{y}}$ and the gas velocity dispersion ($\sigma_{\rm{g}}$) while keeping other parameters fixed to the fiducial model. We keep the boundary conditions fixed in this exercise as they do not matter for the overall profile. We can directly compare these profiles with the oxygen abundance distribution of the youngest stars (ages $< 0.5\,\rm{Gyr}$) in the APOGEE database, for which we can neglect radial migration and assume their birth radii is the same as their current location in the disc. We plot the abundance of these stars as a function of their Galactocentric radius (normalized by $r_0 = 3\,\rm{kpc}$) in \autoref{fig:fig8}. Note that none of the youngest stars lie within $6\,\rm{kpc}$ of the Galaxy center in our sample.

We show the resulting model metallicity profiles in \autoref{fig:fig8}. It is not surprising that the profiles are non-linear, given the complex interplay between various ISM processes that ultimately sets the metallicity in the models (see \autoref{fig:graphical_model}). Such non-linear profiles are routinely observed in the ISMs of external galaxies, even though the gradient reported is based on a linear fit \citep[e.g.,][]{2019A&ARv..27....3M,2019ARA&A..57..511K,2021MNRAS.502.3357P,2023MNRAS.519.4801C}. We first focus on the left panel of \autoref{fig:fig8} which plots model metallicity profiles with all the fiducial model parameters fixed except for $\phi_{\rm{y}}$. We see that model profiles that best encapsulate the extent of the data are those where $\phi_{\rm{y}}$ is neither too high nor too low. Now, we fix $\phi_{\rm{y}} = 0.5$ and vary $\sigma_{\rm{g}}$ from $6\,\rm{km\,s^{-1}}$ (dwarf-galaxy like, \citealt{2013ApJ...773...88S,2015MNRAS.449.3568M,2015AJ....150...47I}) to $40\,\rm{km\,s^{-1}}$ (ULIRG-like, \citealt{2009ApJS..182..628V,2015ApJ...800...70S,2017ApJ...836...66S}). We read off from the right panel of \autoref{fig:fig8} that trends seen in the data cannot be reproduced by models where $\sigma_{\rm{g}}$ is either too low or too high, as expected. For example, the metallicity profile corresponding to $\sigma_{\rm{g}} = 6\,\rm{km\,s^{-1}}$ is inverted, which is not observed in local massive spiral galaxies \citep{2016A&A...587A..70S,2017MNRAS.469..151B,2021MNRAS.502.3357P}. Variations in boundary conditions do not impact these results. The key observable responsible for driving $\phi_{\rm{y}}$ and $\sigma_{\rm{g}}$ to intermediate values for the youngest stars is the cluster of stars at $R_{\rm{gal}} \approx 9 - 12\,\rm{kpc}$ with $\rm{[O/H]} \approx -0.2$. If stars at these distances were more metal rich, our models would suggest a higher $\phi_{\rm{y}}$ (\textit{i.e.,}, an outflow metallicity closer to the ISM metallicity). Extending this analysis to older stars is non trivial because the ISM model metallicity profiles need to be adjusted to take radial migration into account.

\subsection{Degeneracy between model parameters}
\label{s:degeneracy}
Since the metallicity profiles are dictated by the ratios of metal advection, star formation, and gas accretion to metal diffusion ($\mathcal{P},\,\mathcal{S}$, and $\mathcal{A}$, respectively), model parameters that appear together in these ratios are degenerate with each other.\footnote{In principle, one could therefore first constrain $\mathcal{P},\,\mathcal{S}$, and $\mathcal{A}$, and then investigate the combination of model parameters that yield the required values of $\mathcal{P},\,\mathcal{S}$, and $\mathcal{A}$. Such a two-step approach would give identical constraints on the model parameters.} For example, we notice from \autoref{eq:SToomre} that the fractional parameters $f_{\rm{sf}}$ (fraction of star forming gas in the ISM) and $\phi_{\rm{y}}$ (differential enrichment of galactic outflows) only appear in $\mathcal{S}$. These two parameters are thus degenerate -- models where either parameter is very low (\textit{i.e.,} close to zero) cannot explain the observed 2D MDFs, and perform the worst on our metric. If $f_{\rm{sf}}$ is too low or too high, $\phi_{\rm{y}}$ can compensate for its effect on $\mathcal{S}$. As the value of $f_{\rm{sf}}$ is well constrained for Milky Way like galaxies, we have exploited this degeneracy to implicitly place constraints on the value of $\phi_{\rm{y}}$ as above. 

Similarly, the parameters $f_{\rm{gP}}$ (pressure in the disc due to gas self-gravity) and $\phi_{\rm{mp}}$ (ratio of total to the turbulent pressure in the disc) are also degenerate. If we only vary $f_{\rm{gP}}$ independent of $f_{\rm{gQ}}$, we obtain variations in $\mathbb{M}$ in the same direction as when we vary $\phi_{\rm{mp}}$. However, the resulting values of $\mathbb{M}$ do not scale together quantitatively since $\phi_{\rm{mp}}$ is not a fractional parameter. While the overall impact of varying either $f_{\rm{gP}}$ or $\phi_{\rm{mp}}$ is noticeable, they are well constrained by direct observations and analytical calculations, so the degeneracy between the two is not of concern for our analysis.



\section{Conclusions}
\label{s:conclusions}
In this work, we use a combination of state of the art stellar disc model and first principles ISM chemical evolution model to reproduce the observed radius -- metallicity (oxygen abundance) and age -- metallicity distribution of APOGEE red clump stars. We use the stellar, low-$\alpha$ disc model of \cite{2018ApJ...865...96F,2020ApJ...896...15F} to map the present day radius of these stars ($R_{\rm{gal}}$) to their birth radii ($R$), and the ISM chemical evolution model of \cite{2021MNRAS.502.5935S,2023arXiv230315853S} to investigate which ISM parameters are important in reproducing the data. 

Our intention is to provide a qualitative understanding of how we can recover information about the evolution of the Galactic ISM and outflows from stellar data. We do so by keeping the stellar disc model parameters fixed and introducing simple variations in the ISM model parameters, and use different metrics to compare the modeled radius -- metallicity and age -- metallicity distributions with the APOGEE data corrected for selection function effects. The fiducial model that we consider as a starting point for our analysis is inspired from observations of local spiral galaxies. The fiducial model does not reproduce the observations, implying that the ISM history of the Milky Way is somewhat different than a \textit{typical $z=0$ spiral galaxy}.

We first validate models that best reproduce APOGEE data against existing constraints from Galactic as well as extragalactic gas-phase observations. As we discuss in \autoref{s:discussion}, all except one of our best models are in good agreement with numerous gas-phase observations. The exception is the model that allows for strong supernova clustering, and well reproduces APOGEE data, but at the cost of unusually high gas velocity dispersion not observed in the Milky Way or its analogues. This consistency check thus demonstrates that Galactic chemical evolution models should be validated against extragalactic observations as these observations provide a benchmark to compare the Milky Way with, and often contains information about ISM physics not directly accessible within the Galaxy. 

We exploit the successes of the other models best reproducing the APOGEE data to place novel constraints on two key parameters that inform on the evolution of the Galactic ISM and outflows:

\begin{enumerate}
    \item The parameter $\phi_{\rm{y}}$, which describes the differential enrichment of galactic outflows, is less than unity, implying the presence of metal-rich (relative to the local ISM) Galactic outflows over the last 6 Gyr. Our best models constrain $0.4 \leq \phi_{\rm{y}} \leq 0.7$ for the Milky Way, which, assuming a mass loading factor less than unity \citep{2017MNRAS.465.1682H}, implies that the metallicity of outflows is $1.2-1.8\times$ the local ISM metallicity. This finding is also supported by high resolution simulations modeling the ISM -- CGM interface in Solar Neighborhood-like environments \citep{2023arXiv230907955V}. If true, this means that the commonly used assumption that the outflow metallicity equals the ISM metallicity is not applicable to the Milky Way, as is also supported by observations of nearby galaxies \citep{2018MNRAS.481.1690C,2021MNRAS.504...53S}.

    \item The gas velocity dispersion (averaged over the last 6 Gyr) in the Milky Way is higher than that for a galaxy of similar stellar mass, but is within the scatter present in the stellar mass -- velocity dispersion relation \citep{2020MNRAS.495.2265V}. The elevated dispersion can either be sourced from a higher star formation efficiency per freefall time (as is indeed measured for Milky Way molecular clouds -- \citealt{2021ApJ...912L..19P}), increased turbulence due to clustering of supernovae and superbubble formation (which can also enhance outflow metallicity -- \citealt{2020ApJ...903L..34K,2023arXiv230907955V}), or due to gas accretion (expected to be important for massive galaxies like the Milky Way -- \citealt{2022MNRAS.513.6177G}). This highlights the importance of considering energy balance in the ISM in addition to mass and metal balance in chemical evolution models.
\end{enumerate}

Both of these findings are novel because to date, no chemical evolution models have simultaneously included differential enrichment of outflows and energy balance in the ISM. The metal enrichment of outflows (which most influences the match between the ISM models and the data) is the least constrained from both Galactic as well as extragalactic observations \citep[e.g.,][]{2018MNRAS.481.1690C}. We do not fit the model parameters to reproduce the data, and therefore do not expect to reproduce detailed features of the MDF or the radius -- metallicity and age -- metallicity distributions. This will be the subject of a future work involving likelihood-free inference to create a model that specifically applies to the Milky Way. Starting with the Milky Way, this work thus shows the power of combining stellar and gas-phase information in galaxies to reveal their detailed evolutionary history by simultaneously considering both the stars and the ISM. 

\section*{Acknowledgements}
We thank Anthony Brown and the anonymous referee for providing feedback on this manuscript, and Danny Horta, Amina Helmi, and David Weinberg for useful discussions. We also thank Michael J. Greener for their digital transcription of the original review by \cite{1980FCPh....5..287T} that we use in this work. PS is supported by the Leiden University Oort Fellowship and the International Astronomical Union TGF Fellowship. YST acknowledges financial support from the Australian Research Council through the DECRA Fellowship DE220101520. NF is supported by the Natural Sciences and Engineering Research Council of Canada (NSERC) [funding reference number CITA 490888-16] through the CITA postdoctoral fellowship and acknowledges partial support from an Arts \& Sciences Postdoctoral Fellowship at the University of Toronto. This work was performed in part at Aspen Center for Physics, which is supported by National Science Foundation grant PHY-2210452. This work was partially supported by a grant from the Simons Foundation. We acknowledge using the following softwares: Astropy \citep{2013A&A...558A..33A,2018AJ....156..123A,2022ApJ...935..167A}, Numpy \citep{oliphant2006guide,2020arXiv200610256H}, Scipy \citep{2020NatMe..17..261V}, Matplotlib \citep{Hunter:2007}, and cmasher \citep{2020JOSS....5.2004V}.

\section*{Data Availability}
APOGEE DR14 data used in this work is publicly available \href{https://live-sdss4org-dr14.pantheonsite.io/}{here}.



\bibliographystyle{mnras}
\bibliography{references} 


\appendix

\section{Performance of the models on different metrics}
\label{s:app_diffs}
We use four different metrics to compare the model distributions with the data, and use the sum of these metrics to measure the performance of each model. In this section, we investigate whether different metrics provide results that are consistent with each other. We also check if models perform comparably in both the radius -- metallicity and age -- metallicity domains.

\autoref{fig:app_M} plots the value of each metric for all the models (see \autoref{tab:tab2} for a description of the models). From both the panels, we find that on average, $\rm{\mathbb{M}_{EMD}} \approx \rm{\mathbb{M}_{NSR}} \gtrsim \rm{\mathbb{M}_{Bhat}} \approx 10\rm{\mathbb{M}_{KL}}$, albeit with some scatter. Since the KL divergence is a measure of dissimilarity, and NSR and Bhattacharya distance are measures of similarity, the strength of these measures indicates how similar or dissimilar a given model is as compared to the data. Additionally, the variation in the metric value across models for the three dominant metrics remains consistent. This ensures that a single metric does not dominate the overall result, and models where one metric suggests a good match is echoed by other metrics. 

The value of the KL divergence metric is tiny as compared to the other three. This is expected because KL divergence involves the logarithm of the model and data values. Regardless, we see that the trend in the KL divergence metric values closely follows that in the other three metrics - $\rm{\mathbb{M}_{KL}}$ is higher by more than an order of magnitude for models where the other three metrics are also high, which justifies our factor of 10 scaling. This analysis thus ensures that the different metrics are consistent with each other, and at the same time highlights the importance of using multiple metrics to capture the complexity of comparing 2D model and data MDFs.

\begin{figure*}
\includegraphics[width=\textwidth]{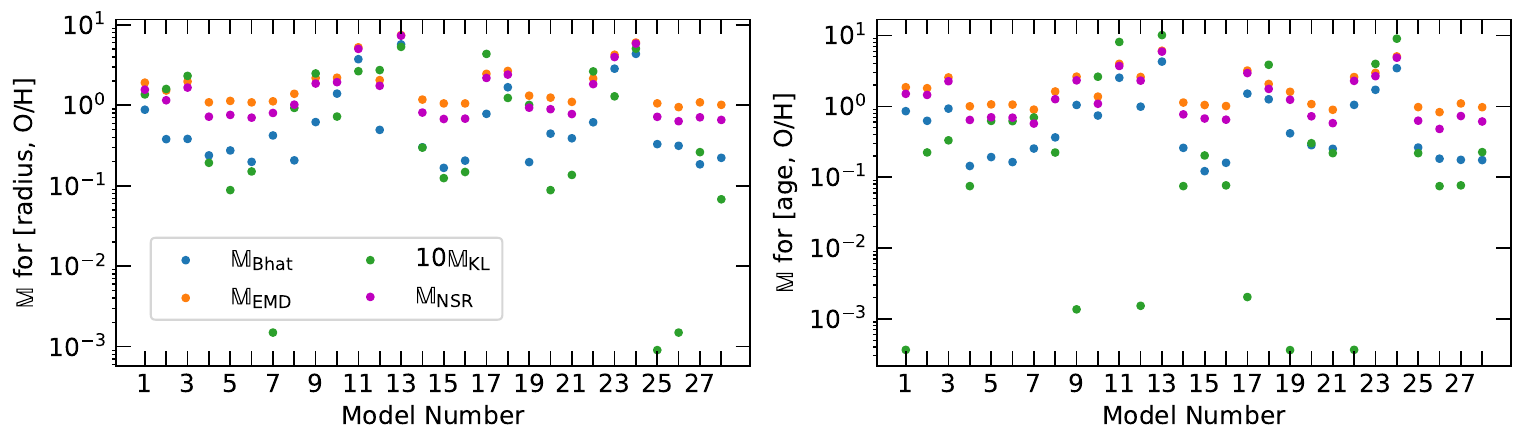}
\caption{\textit{Left panel:} Value of the four different metrics used in the main text (\autoref{s:comparison}) for the radius -- metallicity domain for all the models discussed in this work (see \autoref{tab:tab2}). We scale the KL divergence metric by a factor of 10 so that its mean magnitude is at par with the other metrics. \textit{Right panel:} Same as the left panel but for the age -- metallicity domain. On average, all the metrics yield consistent performance in the context of how a model compares to the data from APOGEE.}
\label{fig:app_blah}
\end{figure*}

\begin{figure}
\includegraphics[width=\columnwidth]{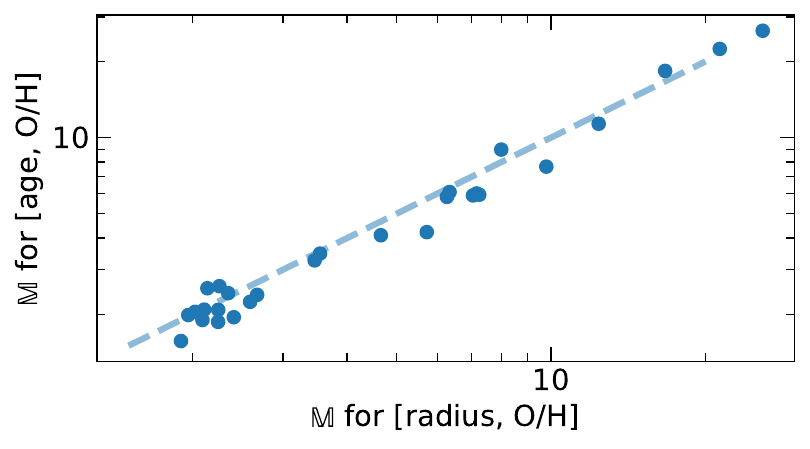}
\caption{Sum of all four metrics in the age -- metallicity domain plotted as a function of that in the radius -- metallicity domain for all the models used in this work. On average, the summed metrics lie close to the 1:1 line, indicating that models that exhibit good performance in one domain also excel in the other domain, and vice versa. This ensures that the result is not biased by one domain over the other.}
\label{fig:app_M}
\end{figure}

\section{Effects of varying the bin size}
\label{s:app_bins}
\begin{figure}
\includegraphics[width=\columnwidth]{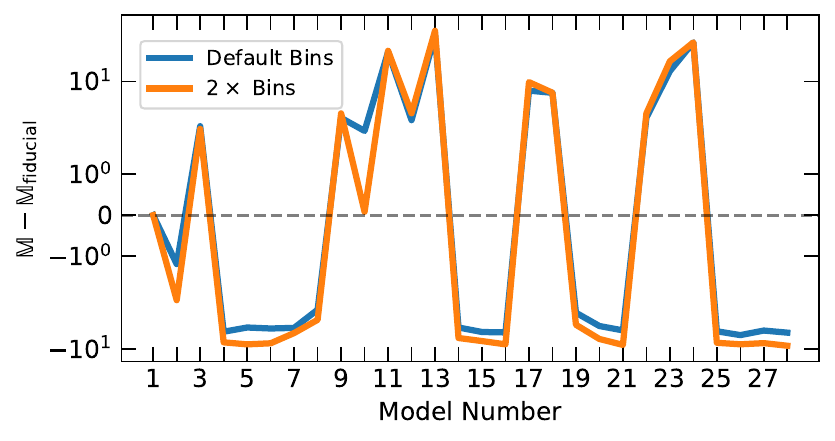}
\caption{Metric value $\mathbb{M}$ for all models relative to the fiducial model, for the default number of bins used in the main text (blue), and for the case where the number of bins is increased by $2\times$ (orange). The relative performance of the models remains the same when more bins are used.}
\label{fig:app_bins}
\end{figure}

In the main text, the bin size in the radius and age dimensions we opt for is inherited from \citet{2020ApJ...896...15F}, which is rather arbitrary and potentially hides substructure at the sub-kpc level in the low-$\alpha$ disc. To test for the effects of the bin size on $\mathbb{M}$, we increase the number of bins in each dimension by a factor of two. \autoref{fig:app_bins} plots $\mathbb{M}$ relative to the $\mathbb{M}$ for the fiducial model. We find that increasing the number of bins has virtually no effect on the relative $\mathbb{M}$, and models that best reproduce the data remain identical between the two cases.

Note that while it is mathematically straightforward to further increase the number of bins to assess the performance of the models, there is a physical limitation for the regime of applicability of the underlying chemical evolution model: the ISM component is designed to explain chemical enrichment on $\sim\rm{kpc}$ scales and does not include additional physics, azimuthal variations \citep{2018A&A...618A..64H,2019ApJ...887...80K}, and stochasticity that acts on smaller scales \citep{2018MNRAS.475.2236K,2024MNRAS.529..104M}, so using bin sizes $\lesssim 300\,\rm{pc}$ will not yield a physically meaningful result.

\section{Radius -- metallicity and age -- metallicity distributions from all other models}
\label{s:app_MDFs}
\begin{figure*}
\includegraphics[width=0.49\textwidth]{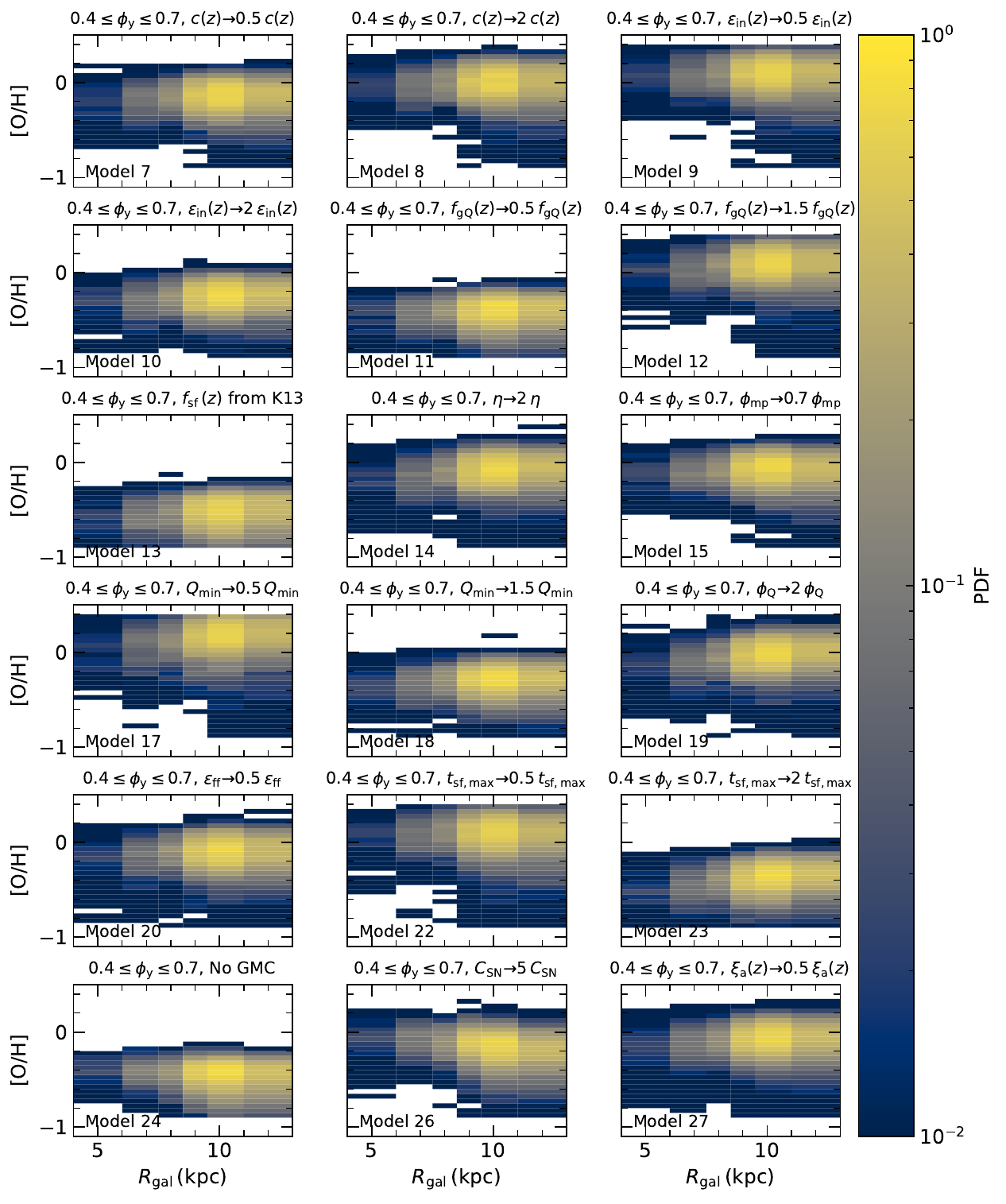}
\includegraphics[width=0.49\textwidth]{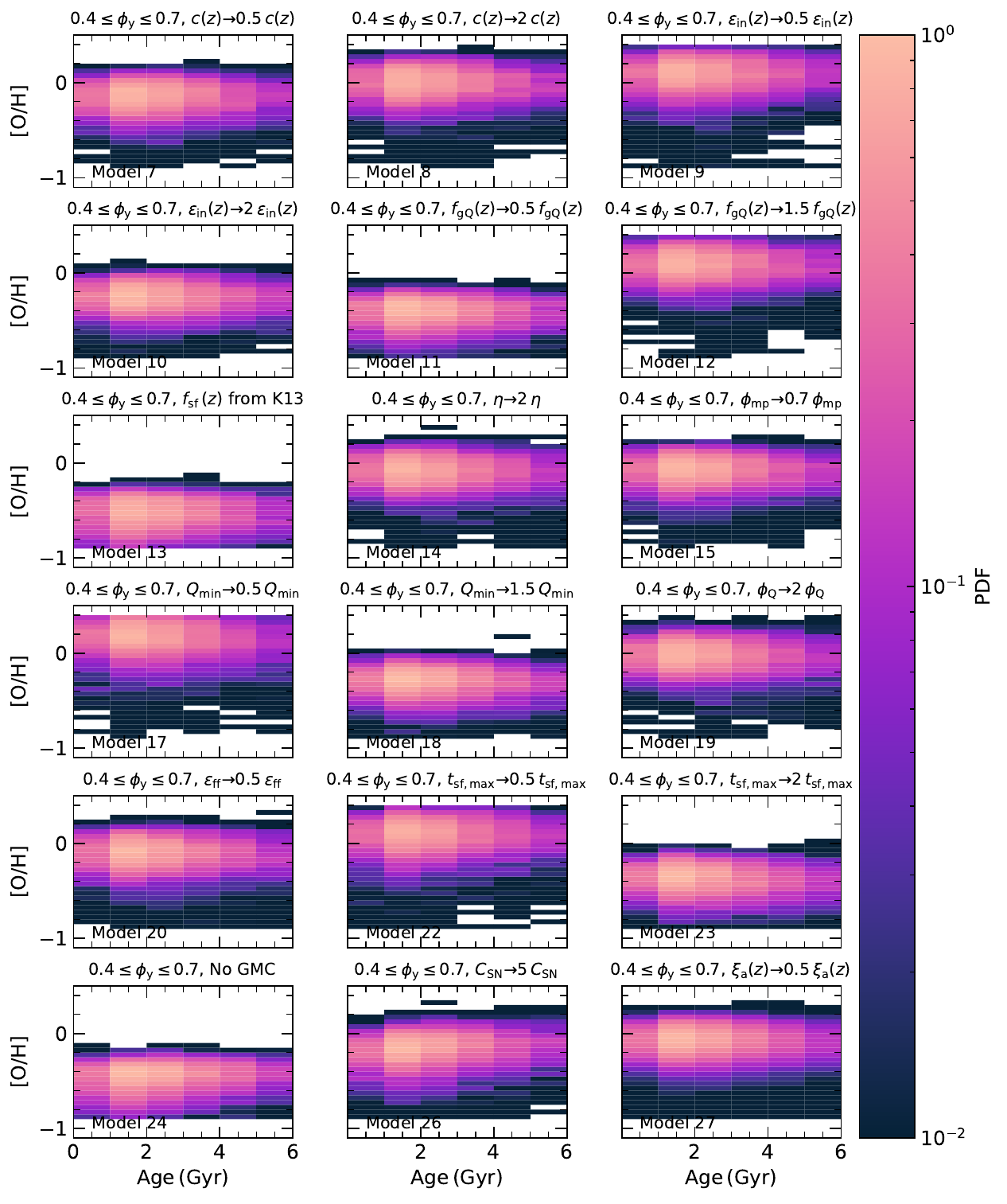}
\caption{Radius -- metallicity and age -- metallicity distributions of all other models not shown in the main text. Titles on each subplot denote the variation introduced to create the respective model (see \autoref{tab:tab2} for details).}
\label{fig:allothers}
\end{figure*}

\autoref{fig:allothers} plots the radius -- metallicity and age -- metallicity distributions for all the other models we discuss in the main text. These models perform poorly as compared to models presented in \autoref{fig:fig5}.

\bsp	
\label{lastpage}
\end{document}